\documentclass[aps,prl,twocolumn,superscriptaddress,groupedaddress]{revtex4}  
\usepackage{graphicx}  
\usepackage{dcolumn}   
\usepackage{bm}        
\usepackage{amssymb}   
\usepackage{subfigure}
\usepackage{multirow}

\usepackage{natbib}
\usepackage{float}
\usepackage{slashed}
\usepackage{amsmath}
\usepackage{accents}
\newlength{\dhatheight}

\newcommand{\BR}{\mathcal{B}}
\begin{document}



\title{Search for Flavor Changing Neutral Currents in $t\to H c, H\to \tau\tau$ Decay at the LHC}
\author{ Xin Chen}\affiliation{Department of Physics, Tsinghua University, Beijing 100084, China}
\affiliation{Collaborative Innovation Center of Quantum Matter, Beijing 100084, China}
\author{ Li-Gang Xia}\affiliation{Department of Physics, Tsinghua University, Beijing 100084, China}
                             

\begin{abstract}

The prospects of searching for the flavor changing neutral current effect in the decay of $t\to H c, H\to \tau\tau$ are investigated with the simulated  $p-p$ collision data for the ATLAS detector at the LHC, where the Higgs mass is assumed to be 125 GeV. A fit based on the constraints from the Higgs mass and the tau decay kinematics is performed for each event, which improves significantly the Higgs and top mass reconstruction and helps the signal-background separation. Boosted Decision Trees discriminants are developed to achieve an optimal sensitivity of searching for the FCNC signal. An expected upper limit of the branching ratio $\BR(t\to H c)$  at $95\%$ confidence level of 0.25\% is obtained with a data set of 100~fb$^{-1}$ at $\sqrt{s}=13$ TeV during the LHC Run-2 period.
\end{abstract}

\maketitle

\section{Introduction}
In the Standard Model (SM), the flavour-changing neutral current (FCNC) interactions are extremely suppressed due to the Glashow-Iliopoulos-Maiani mechanism~\cite{GIM}. An enhancement of such FCNC processes may provide indirect evidence of new physics beyond the SM.  The FCNC process has been searched for in the flavor physics such as $B_s\to \mu\mu$~\cite{LHCb_Bs} and in the vector boson mediated process of $t\to Zq$~\cite{tZq}.  After the discovery of the Higgs boson~\cite{higgs_atlas, higgs_cms} at the Large Hadron Collider (LHC), it is possible that the new scalar could be connected to the new physics. Searching for Higgs-related FCNC processes is an important way to study the Higgs' properties. In this context, the FCNC process of $t \to Hc$ becomes more interesting since the Higgs boson, with a measured mass $m_H = 125.09\pm0.24$ GeV~\cite{higgs_mass}, is lighter than the top quark. The large mass of the $t$ and $c$-quark could significantly enhance the $t \to Hc$  decay probability, leading to a measurable rate~\cite{cheng_sher, he_1, he_2}. 

The branching ratio of $t \to Hc$, $\BR(t \to Hc)$, is predicted to be of the order of $10^{-15}$~\cite{brtch1, brtch2, brtch3, brtch4} in the SM. This FCNC decay is enhanced in various SM extentions. For example, the two Higgs-doublet models (2HDM) predict that $\BR(t\to Hc)$ ranges $10^{-5}$ up to $10^{-3}$~\cite{brtch5, brtch6, brtch7, brtch8, brtch9, brtch10, brtch11}, which is already close to the measured upper limits, 0.56\%~\cite{fcnc_cms} and 0.46\%~\cite{fcnc_atlas} from the CMS and ATLAS experiments respectively. Compared to the decay modes of  $H\to \gamma\gamma$ and $H \to b\bar{b}$, the Higgs mass can not be easily reconstructed in the decay mode $H \to \tau\tau$, because of the presence of the neutrinos. Usually, the collinear approximation~\cite{colapprox1, colapprox2, colapprox3} or the Missing Mass Calculator (MMC) technique~\cite{MMC} is used. In this paper, we propose a method to improve the sensitivity for the decay mode $H \to \tau\tau$ based on the kinematic properties of the production $pp \to t\bar{t} \to WbHc$ with $W\to jj$ at the LHC. This result can be combined with the existing search modes for the best sensitivity. It can confirm the signal as a FCNC type in the case of discovery as the $H\to \gamma\gamma$ does, and it does not suffer from the $b$-jet combinatorics as in the $H \to b\bar{b}$ channel.

\section{Background analysis and MC production}
Assuming a CP-even Higgs  according to the LHC measurements \cite{higgs_cp}, the following effective Lagrangian term is added to study the FCNC property of the Higgs boson.
\begin{equation}
\mathcal{L} = \lambda_{tcH} \bar{c}t H + h.c.,
\label{eq:eq1}
\end{equation}
where $\lambda_{tcH}$ is the coupling constant.
The major SM backgrounds for the $H\to \tau\tau$ channel are $t\bar{t}$ (+ 0/1/2 jets), single top ($tW^-+c.c.$) and $Z/\gamma^{\star}\to ee/\mu\mu/\tau\tau$ + heavy flavor jets ($c\bar{c}/b\bar{b}$ + 0/1/2 jets). The $t\bar{t}$ cross section at 13 TeV is normalized to $820$ pb at next-to-next-to-leading-order (NNLO) for a top mass of 173 GeV~\cite{ttbar_xs}. The top quark, $W$ boson and $\tau$ lepton decay inclusively. Other backgrounds, such as the multijet background, are significantly reduced by requiring a $b$-tagged jet and two $\tau$'s, and are not considered. The $t\bar{t}+W/Z/H$ backgrounds are less than $1\%$ of the top backgrounds. Therefore they are not considered as well. The signal shares the same production with the $t\bar{t}$ events, except that one top quark ($t$ or $\bar{t}$) decays to $H+c/\bar{c}$, and the other decays into hadronic jets ($t\to Wb\to jjb$). This is different from the FCNC search with the di-$\tau$ channel in Refs.~\cite{fcnc_atlas, fcnc_pheno}, in which the other top decays leptonically and the Higgs mass could not be reconstructed. In this work, both the signal and background samples are generated with MadGraph5 \cite{MG5}, the parton shower is provided by Pythia \cite{Pythia}, and the detector response is simulated by Delphes \cite{Delphes}. In the Delphes simulation, the leptons ($e/\mu$) have combined tracking, reconstruction and identification efficiencies of 70-95\% depending on the transverse momentum $p_{\text{T}}$ and pseudorapidity $\eta$. They are required to have a minimum $p_{\text{T}}$ of 15 GeV. As in the default settings in~\cite{Delphes}, the $b$-tagging efficiency for real $b$-jets ($c$-jets) is 40-50\% (10-20\%) in the high $p_{\text{T}}$ region, with a flat fake efficiency of 0.1\% for all other light jets~\cite{BTag}. No explicit $c$-tagging is used, as it is very hard to make a good separation between a $c$-jet and a $b$-jet. The hadronic $\tau$ identification efficiency is 40\% for real $\tau$'s, and is 1\% for jets faking the $\tau$~\cite{TauPerf}. They should have a minimum  $p_{\text{T}}$ of 25 GeV. The anti-$k_t$ algorithm with a radius parameter 0.4 is used to cluster jets. These parameters are generally in accordance with the expected performance of the ATLAS detector during LHC Run-1 period, based on which the current analysis is carried out.
\section{Event Selection}
In the initial event selection, exactly two tau candidates ($e$, $\mu$, and the hadronic $\tau$-jet) with opposite charge sign are required. To comply with the triggers, in the $\tau_l\tau_l$ and $\tau_l\tau_h$ channels, the leading $e/\mu$ should have $p_{\text{T}}>20$ GeV. Here $\tau_l$ ($\tau_h$) denotes the leptonic (hadronic) decay of the tau lepton. In the $\tau_h\tau_h$ channel, the leading (subleading) $\tau_h$ should have $p_{\text{T}}>40$ GeV ($p_{\text{T}}>30$ GeV).  The leptons, $\tau$ and $b$-jets are only defined in the tracking volume with $|\eta|<2.5$. 

\begin{figure}[htb]
\centering
\subfigure{\includegraphics[width=0.4\textwidth]{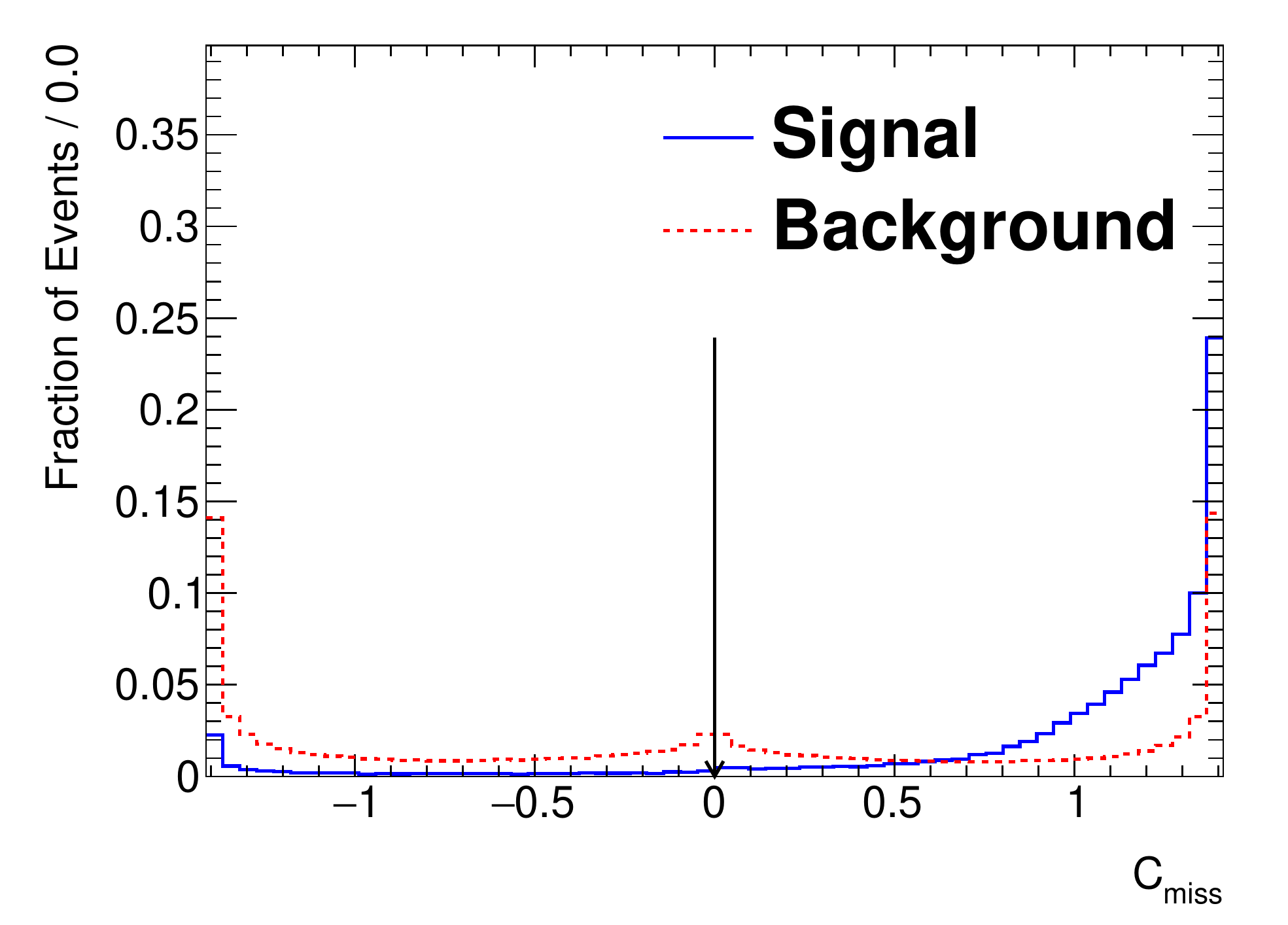}}
\caption{\label{fig:MET_cent} The distributions of the $\slashed{E}_{\text{T}}$ centrality, $C_{\text{miss}}$, in the signal (blue-solid histogram) and in the background (red-dashed histogram).  The arrow represent the cut used in the event selection.}
\end{figure}

The missing transverse energy, $\slashed{E}_{\text{T}}$, is required to be larger than 20~GeV. To suppress the background, it is further required that the $\slashed{E}_{\text{T}}$ centrality, $C_{\text{miss}}$, is greater than zero as shown in Fig.~\ref{fig:MET_cent}. The $C_{\text{miss}}$ is defined as
\begin{eqnarray}
\begin{array}{l}
C_{\text{miss}} = {(x+y)}/{\sqrt{x^2+y^2}}, \\
\text{with}~x=\frac{\sin(\phi_{\text{mis}}-\phi_{l1})}{\sin(\phi_{l2}-\phi_{l1})},  y=\frac{\sin(\phi_{l2}-\phi_{\text{mis}})}{\sin(\phi_{l2}-\phi_{l1})} ,
\end{array}
\label{eq:eq3}
\end{eqnarray}
where $\phi_{l1,2}$ are the azimuthal angle of the two leptons ($e,\mu$, or $\tau_h$) in the transverse plane, and $\phi_{\text{mis}}$ is the azimuthal angle of the missing energy.

\begin{figure}[htb]
\centering
\subfigure{\includegraphics[width=0.4\textwidth]{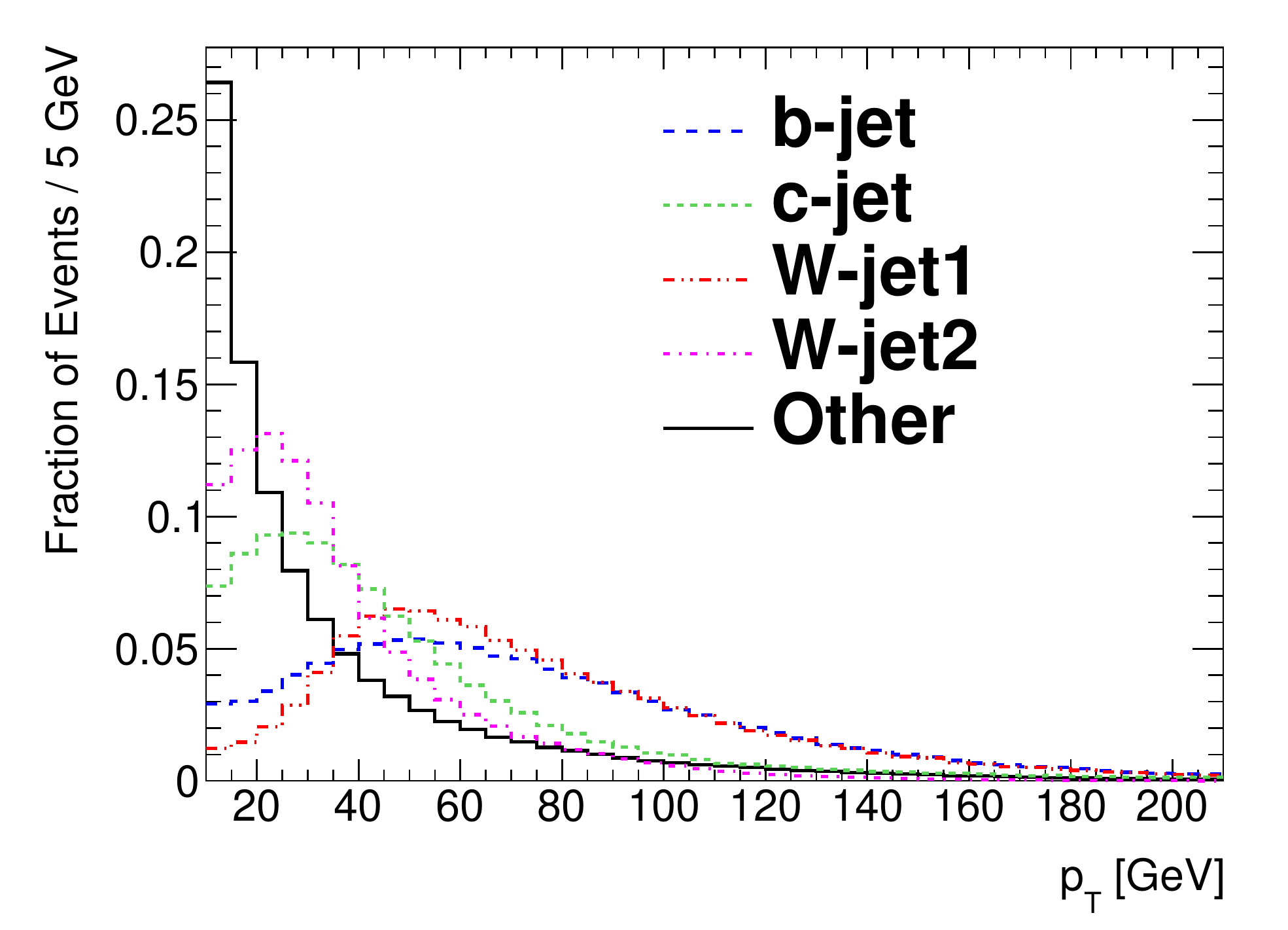}}
\caption{\label{fig:jetPt} The $p_{\text{T}}$ distributions of the jets in the signal at truth level. $W$-jet1 ($W$-jet2) denotes the leading (subleading) jet from the $W$ decay. ``Other'' refers to the jets not matched to any parton in the top decay. }
\end{figure}

Figure~\ref{fig:jetPt} shows the distributions of the transverse momentum, $p_{\text{T}}$, of the involved jets in the signal at the truth level. At the reconstruction level, all jets should have $p_{\text{T}} > 30$ GeV with $|\eta|<4.5$. They are required to not overlap with the leptons and $\tau$'s. The energy and $p_{\text{T}}$ of all objects are smeared according to their experimental resolutions. 

\begin{figure}[htb]
\centering
\subfigure{\includegraphics[width=0.4\textwidth]{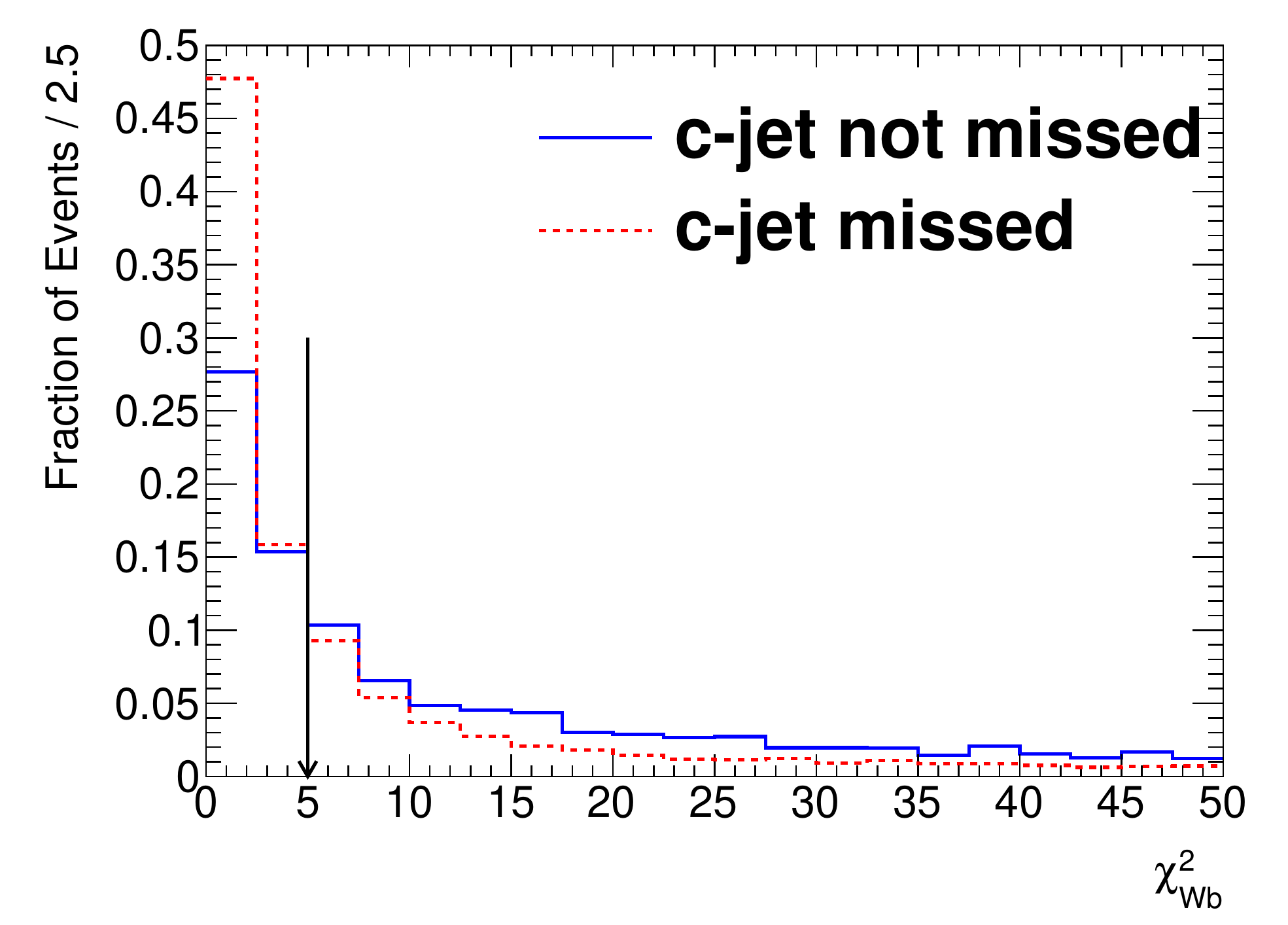}}
\caption{\label{fig:chi2_wb} The distributions of $\chi^2_{Wb}$  in the 3-jet events. The blue-solid (red-dashed) histogram represents the distributions of the events with the $c$-jet found (missing) in the signal.  The arrow represents the cut used in the event selection.}
\end{figure}

\begin{figure*}[htb]
\centering
\subfigure{\includegraphics[width=0.4\textwidth]{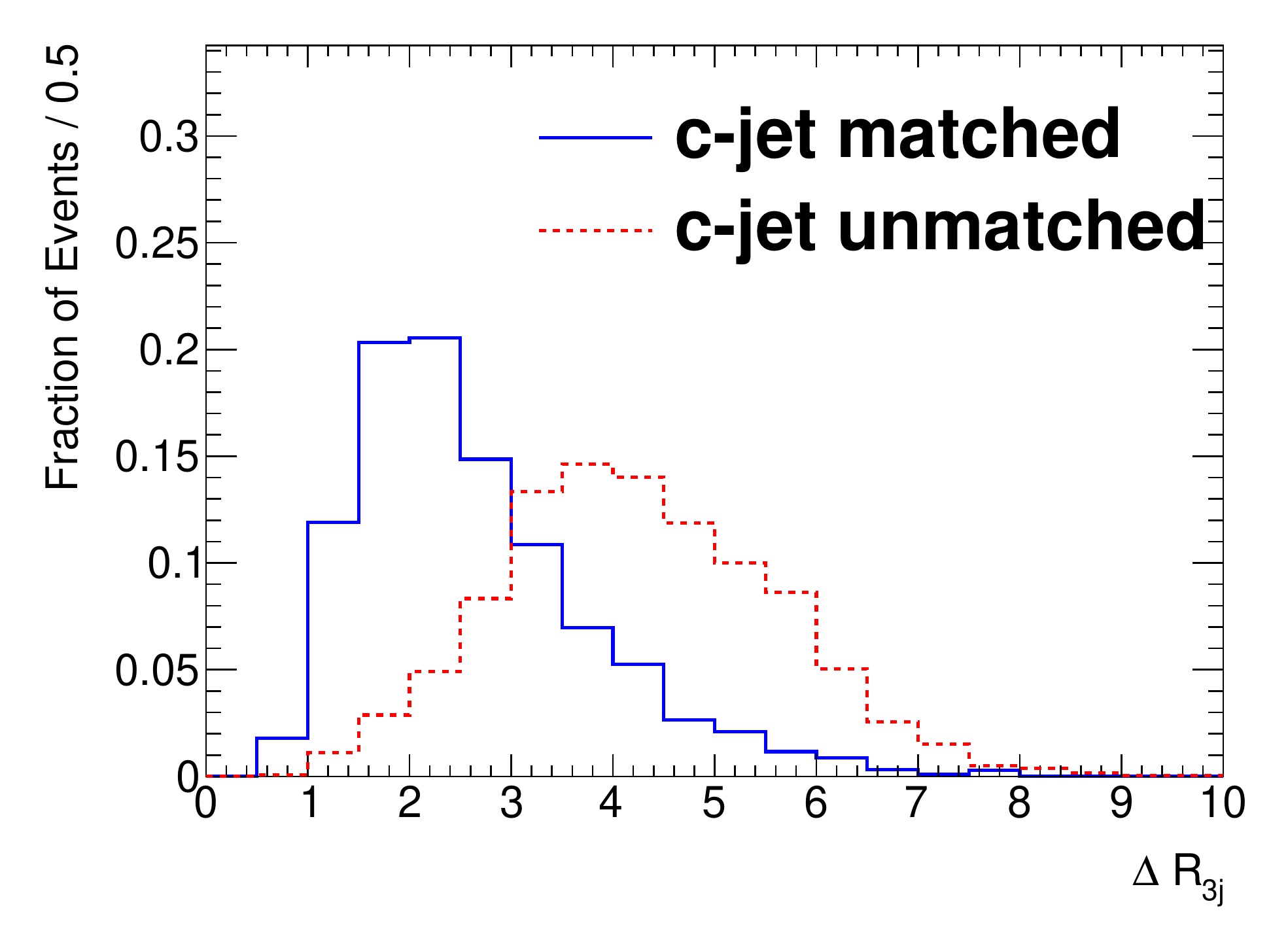}}
\put(-160, 120){\textbf{(a)}}
\subfigure{\includegraphics[width=0.4\textwidth]{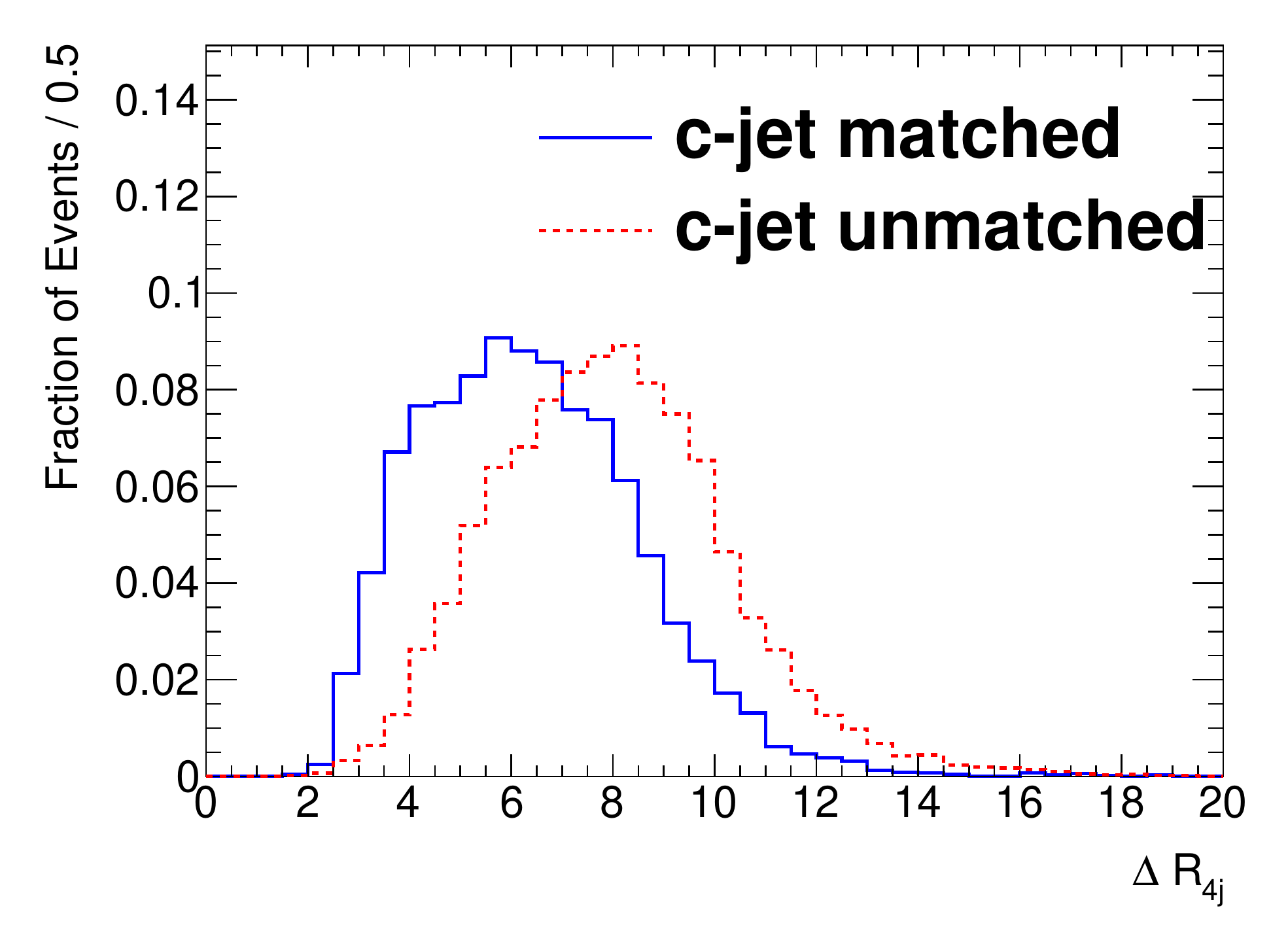}}
\put(-160, 120){\textbf{(b)}}\\
\caption{\label{fig:dR_3j4j} The distributions of $\Delta R_{3j}$ (a) for the 3-jet events and $\Delta R_{4j}$ (b) for the events with at least four jets in the signal. The blue-solid (red-dashed) histograms represent the distributions with the $c$-jet matched (not matched) to the $c$-parton. }
\end{figure*}

To comply to the signal topology, in each event, at least one jet should be tagged as a $b$-jet. If more than one jet is $b$-tagged, the one with the highest $p_{\text{T}}$ is identified as the $b$-jet candidate. If all jets from the top hadronic decay and the $c$-jet from $t\to Hc$ pass the jet selection, there should be at least four jets. However, as can be seen from Fig.~\ref{fig:jetPt}, there are chances that some jets have $p_{\text{T}}$ less than 30 GeV and  may fail to pass the selection. The most likely missing jet is the subleading jet from $W$ decay. These 3-jet events are still kept if the $c$-jet can be found and matched with the Higgs to reconstruct the top. It is done as follows. In the 3-jet events, if the three jets, denoted by $j_1$, $j_2$, $b$, satisfy
\begin{equation}
\chi_{Wb}^2 \equiv \left(\frac{m_{j_1 j_2}-80}{20}\right)^2 + \left(\frac{m_{j_1 j_2 b}-173}{25}\right)^2 <5,
\label{eq:eq4}
\end{equation}
where the mass is in GeV, the event is discarded, as indicated in Fig.~\ref{fig:chi2_wb}. In these events, a good hadronic top is reconstructed, but the $c$-jet from the other top is missing.
If Eq.~(\ref{eq:eq4}) is not satisfied, the $c$-jet is identified by the least sum of angular distances, $\Delta R_{3j}$, as indicated in Fig.~\ref{fig:dR_3j4j}(a). 
\begin{equation}
\Delta R_{3j} \equiv \Delta R(j_c ,H) + \Delta R(j_W ,b) \: .
\label{eq:eq5}
\end{equation}
Here $j_W$ is the jet from $W$ decay. $\Delta R(j_1,j_2)$ is the angular distance of two objects defined as $\Delta R(j_1,j_2)=\sqrt{(\eta_{j_1}-\eta_{j_2})^2 + (\phi_{j_1}-\phi_{j_2})^2}$, with $\phi$ being the azimuthal angle.  

For the events with at least four jets (denoted as 4-jet events), the three leading ones other than the $b$-jet are considered. Out of the three possible combinations that form the two sets of top decay products, the one with the least sum of angular distances, $\Delta R_{4j}$, is chosen, as shown in Fig.~\ref{fig:dR_3j4j}(b). The sum is defined as
\begin{equation}
\Delta R_{4j} \equiv \Delta R(j_c ,H) + \Delta R(j_{1} ,b) + \Delta R(j_{2} ,b) + \Delta R(j_{1} ,j_{2}),
\label{eq:eq6}
\end{equation}
where $j_{1,2}$ are the jets from the $W$ decay. No explicit $c$-tagging is used, and as can be seen above, the $c$-jet is found through pure kinematic criteria such as $\Delta R$.

In order not to dilute the signal extraction, and considering the relatively high fake $\tau$ rate and multiple jets in the signal events, the hadronic $\tau$ jet is required to be truth-matched in the signal, and the truth-matching efficiency (being $87\%$, $78\%$ for $\tau_l\tau_h$, $\tau_h\tau_h$ respectively) is included in the signal acceptance. If either of the $\tau$'s is mis-matched, both the Higgs mass and the top mass will not be reconstructed correctly, which makes the signal events background-like.

\begin{figure}[htb]
\centering
\subfigure{\includegraphics[width=0.4\textwidth]{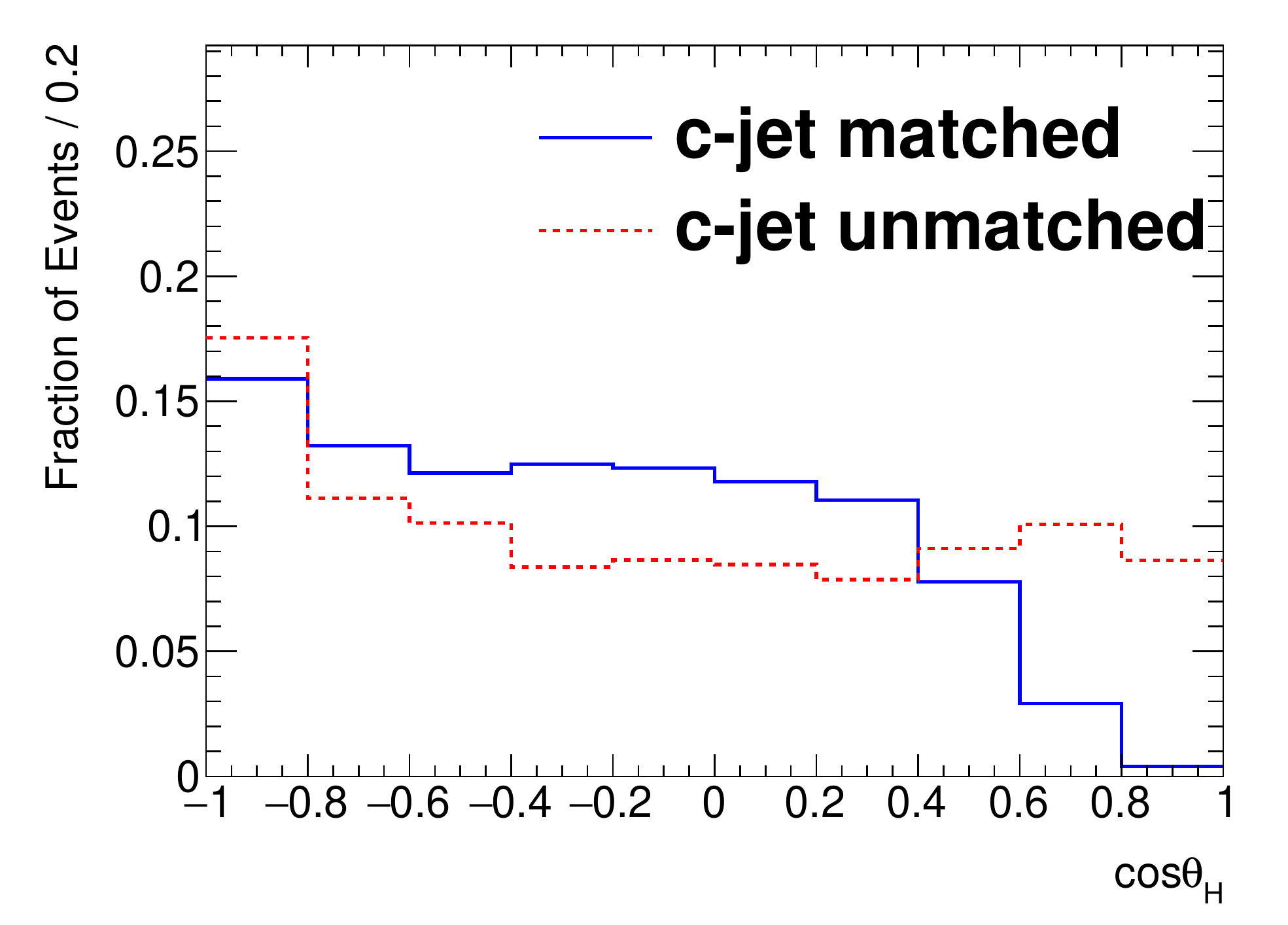}}
\caption{\label{fig:cos_h} The distributions of the Higgs helicity angle, $\cos\theta_H$, in the signal before the $c$-jet matching rate improvement. The blue-solid (red-dashed) histogram represents the distribution with the $c$-jet matched (not matched) to the $c$-parton. }
\end{figure}

To improve the rate of $c$-jet matching with the $c$-parton, the helicity angle of the Higgs, $\theta_H$, is studied. For the decay $t\to Hc$, $\theta_H$ is defined as the angle between the momentum of the top quark in the laboratory frame and the momentum of the Higgs in the center-of-mass frame of the top quark. Fig.~\ref{fig:cos_h} shows the distributions of $\cos\theta_H$ for the events with $m_{bj_1}>170$ GeV, where $m_{bj_1}$ is the invariant mass of the $b$-jet and the leading jet presumably from the $W$ decay. Compared to the distribution of $\cos\theta_H$ with the $c$-jet correctly matched to the $c$-parton, the excess around $\pm 1$ in the falsely matched events is due to the $b$-jet being mis-identified as the $c$-jet candidate. Therefore, if the Higgs helicity angle satisfies $\cos\theta_H<-0.8$ or $\cos\theta_H>0.5$ for the events with $m_{bj_1}>170$ GeV, the $b$-jet candidate and the $c$-jet candidate are exchanged. This improves the $c$-jet matching rate by $2\%$.

After these steps, the final fraction of signal events with the $c$-jet matched to the $c$-parton is found to be around $37\%$.

With the selection conditions described above applied, the expected numbers of the signal and background events for a data set of 10~fb$^{-1}$ in different signal regions are summarized in Tab.~\ref{tab:tab2}.
\begin{table}
\caption{The expected number of signal and background events for a data set of 10~fb$^{-1}$, assuming $\BR(t\to Hc)=1\%$.}
\begin{ruledtabular}
\begin{tabular}{lcccccc}
 & \multicolumn{2}{c}{$\tau_l\tau_l$} & \multicolumn{2}{c}{$\tau_l\tau_h$} & \multicolumn{2}{c}{$\tau_h\tau_h$} \\ \hline
 & 3-jet & 4-jet & 3-jet & 4-jet & 3-jet & 4-jet \\ \hline
$t\bar{t}$ & 6501 & 6411 & 2640 & 4469 & 230 & 561 \\
$Z/\gamma^{\star}$ & 324 & 262 & 42 & 31 & 14 & 11 \\
single top & 220 & 76 & 132 & 99 & 17 & 12 \\
signal & 4.2 & 9.7 & 8.3 & 19.8 & 3.9 & 9.5 \\
\end{tabular}
\label{tab:tab2}
\end{ruledtabular}
\end{table}

\section{Kinematic Constraints}
Since the $\tau$'s in this study are highly boosted objects, the neutrino(s) from the tau decay tends to be aligned with the visible decay products (charged leptons or hadronic $\tau$-jet). Assuming that the visible and invisible parts are collinear, the invisible momentum for each of the two $\tau$'s can be uniquely determined using the measured $\slashed{E}_{\text{T}}$. Thus the invariant mass of the tau lepton pair, $m_H^{\text{col}}$, can be reconstructed with an efficiency loss~\cite{colapprox1, colapprox2, colapprox3}. However, with the MMC which is widely used in the ATLAS analyses involving $\tau$'s, one can have a better di-$\tau$ mass reconstruction by taking the $\tau$ decay kinematics into account. 

In light of the idea in the MMC, the probability distribution of the angular distance of the visible and invisible decay products in the tau decay, denoted by $\mathcal{P}(\Delta R)$, can be parametrized as a function of the transverse momentum of the tau lepton. In the $\tau_l$ mode where two neutrinos are present, it is extended to be the joint probability distribution of $\Delta R$ and $m_{\text{mis}}$ with $m_{\text{mis}}$ being the invariant mass of the neutrinos, denoted by $\mathcal{P}(\Delta R, m_{\text{mis}})$. The probability density functions, $\mathcal{P}(\Delta R)$ and $\mathcal{P}(\Delta R, m_{\text{mis}})$, are obtained from the Monte Carlo (MC) simulation. 
To determine the 4-momenta of the invisible decay products of the tau decays, we reconstruct the following $\chi^2$ in Eq.~\ref{eq:eq2}, based on the probability functions above and the constraints from the tau mass, the Higgs mass and the measured $\slashed{E}_{\text{T}}$.
%
\begin{eqnarray}
\begin{array}{ll}
\chi^2 = & -2\ln \mathcal{P}_1 -2\ln \mathcal{P}_2 + \left( \frac{m_{\tau_1}^{\text{fit}} - 1.78}{\sigma_{\tau}} \right)^2 + \\ 
 & \left( \frac{m_{\tau_2}^{\text{fit}} - 1.78}{\sigma_{\tau}} \right)^2 +  \left( \frac{m_{H}^{\text{fit}} - 125}{\sigma_{\text{Higgs}}} \right)^2 + \\
 & \left( \frac{\slashed{E}_{x}^{\text{fit}} - \slashed{E}_{x}}{\sigma_{\text{mis}}} \right)^2 + \left( \frac{\slashed{E}_{y}^{\text{fit}} - \slashed{E}_{y}}{\sigma_{\text{mis}}} \right)^2 .
\end{array}
\label{eq:eq2}
\end{eqnarray}
Here, the free parameters scanned are the 4-momentum components of the invisible decay products for each tau decay. In the $\tau_h$ mode, only three momentum components are scanned since a single neutrino is massless. $m_{\tau_{1,2}}^{\text{fit}}$,  $m_{H}^{\text{fit}}$ and $\slashed{E}_{x,y}^{\text{fit}}$ are the calculated tau mass, Higgs mass, and missing transverse energy with the scanned parameters. The corresponding mass resolutions, $\sigma_{\tau}$ and $\sigma_{\text{Higgs}}$, are set to 1.8~GeV and 20~GeV respectively. The $\slashed{E}_{\text{T}}$ resolution is taken to be $\sigma_{\text{mis}}=0.53\sqrt{\Sigma E_\text{T}}$ ($\Sigma E_\text{T}$, defined in GeV, is the $E_\text{T}$ sum of all visible objects in an event). The invisible 4-momenta are obtained by minimizing the combined $\chi^2$ for each event. Compared to Ref.~\cite{MMC}, the $3^\text{th}$ to $5^\text{th}$ terms in Eq.~\ref{eq:eq2} are new in our analysis. In~\cite{MMC}, the two $\tau$ mass constraint terms are used when solving for the other unknowns. In our approach, we allow resolutions of the $\tau$ mass terms, taking into account the resolutions on the $\tau$ visible energy measurements. By adding the Higgs mass constraint term in the kinematic fit, not only is the Higgs mass resolution improved, but also the resolutions of the Higgs boson's four-momentum, and the mass of the top from which the Higgs comes.

\begin{figure*}[htb]
\centering
\subfigure{\includegraphics[width=0.33\textwidth]{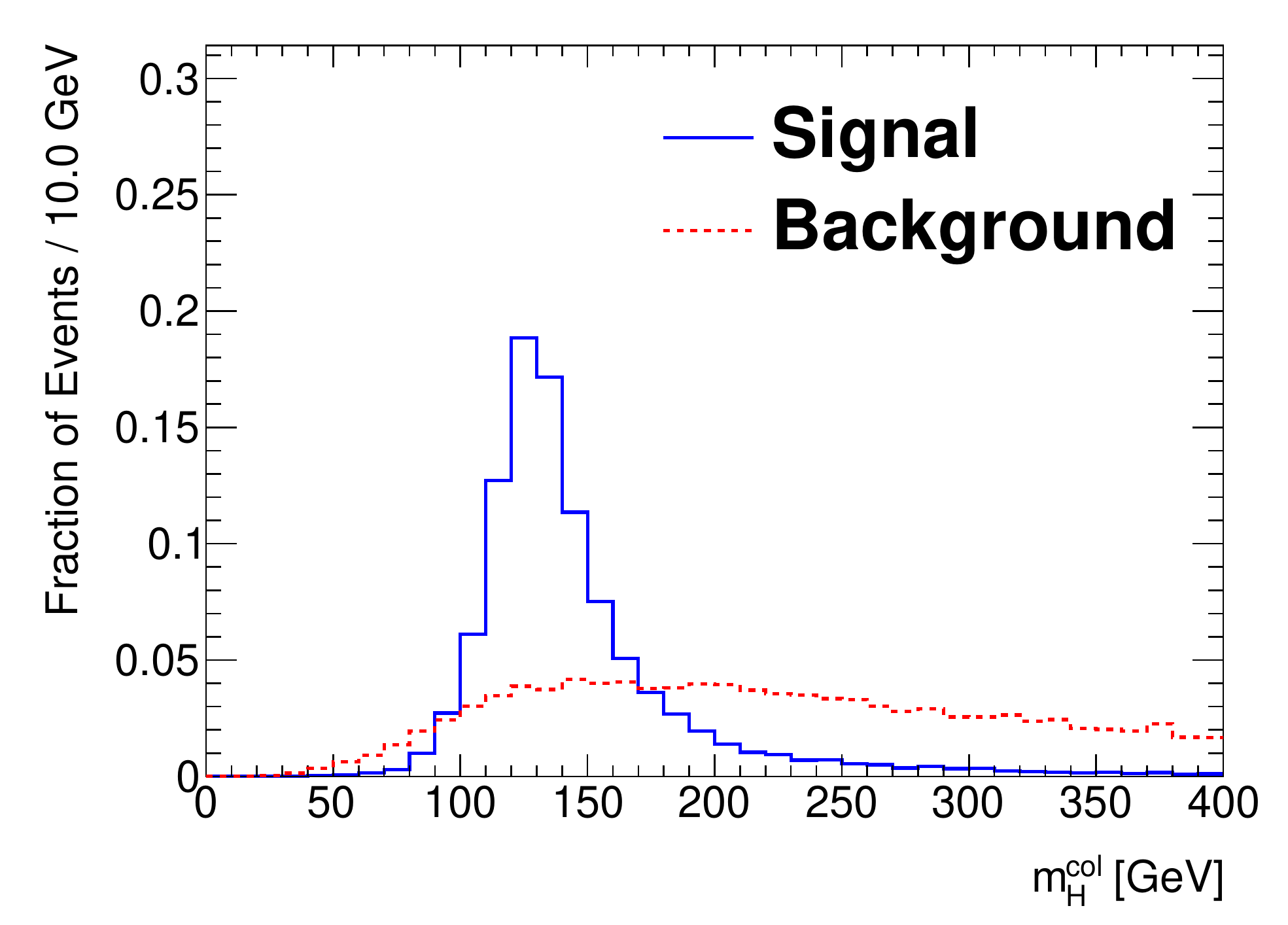}}
\put(-130, 100){\textbf{(a)}}
\subfigure{\includegraphics[width=0.33\textwidth]{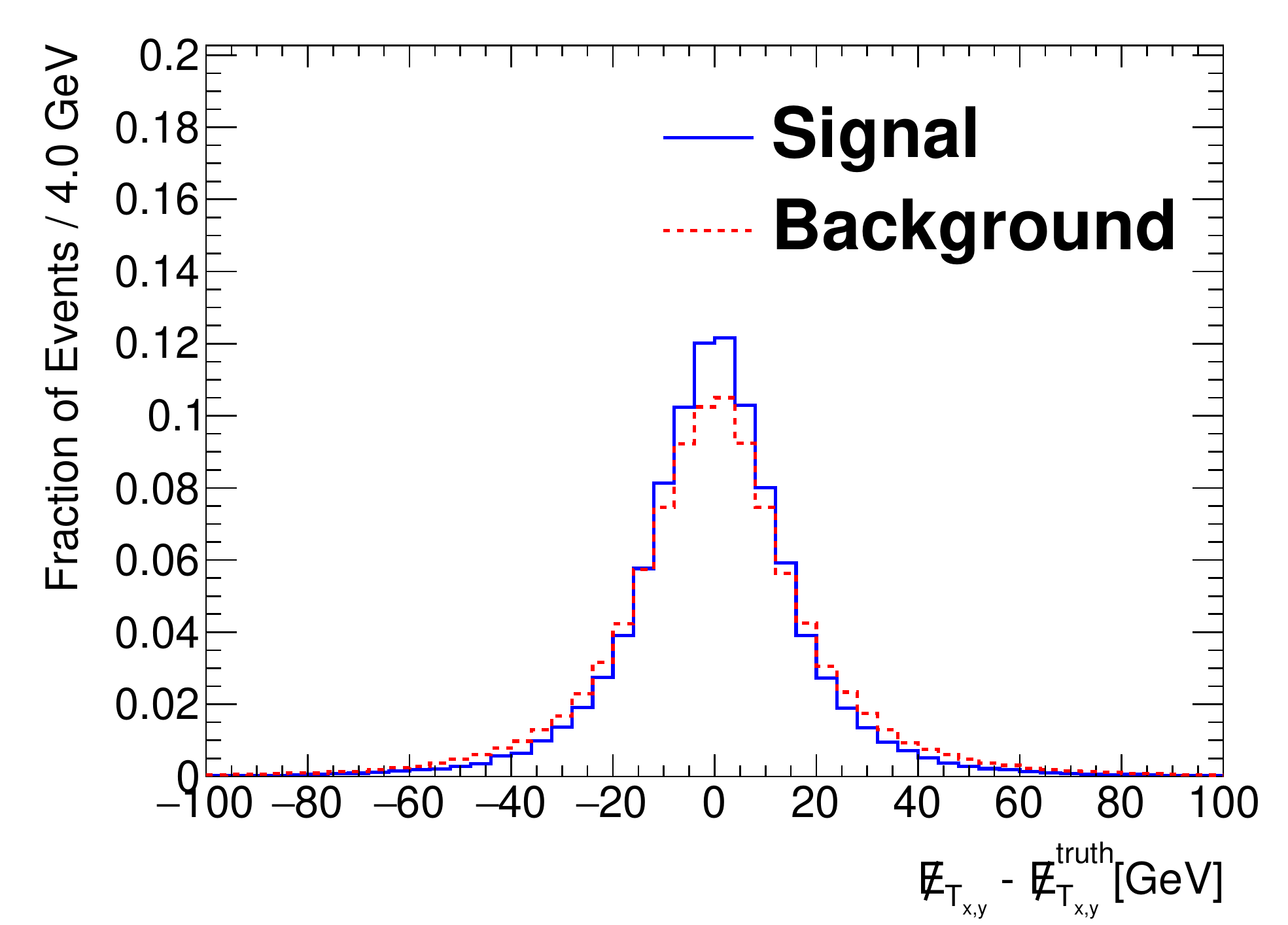}}
\put(-130, 100){\textbf{(c)}}
\subfigure{\includegraphics[width=0.33\textwidth]{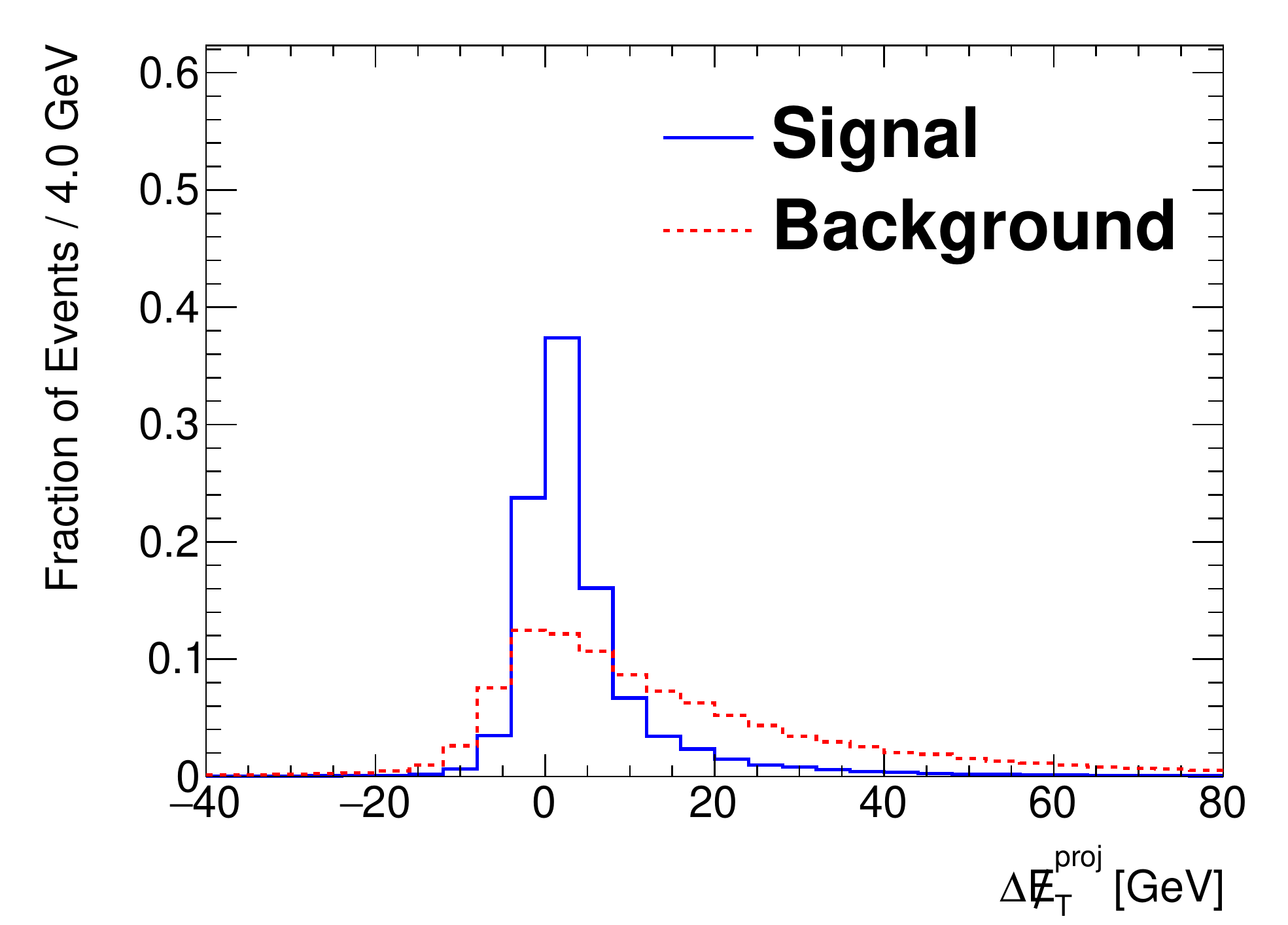}}
\put(-130, 100){\textbf{(e)}}\\
\subfigure{\includegraphics[width=0.33\textwidth]{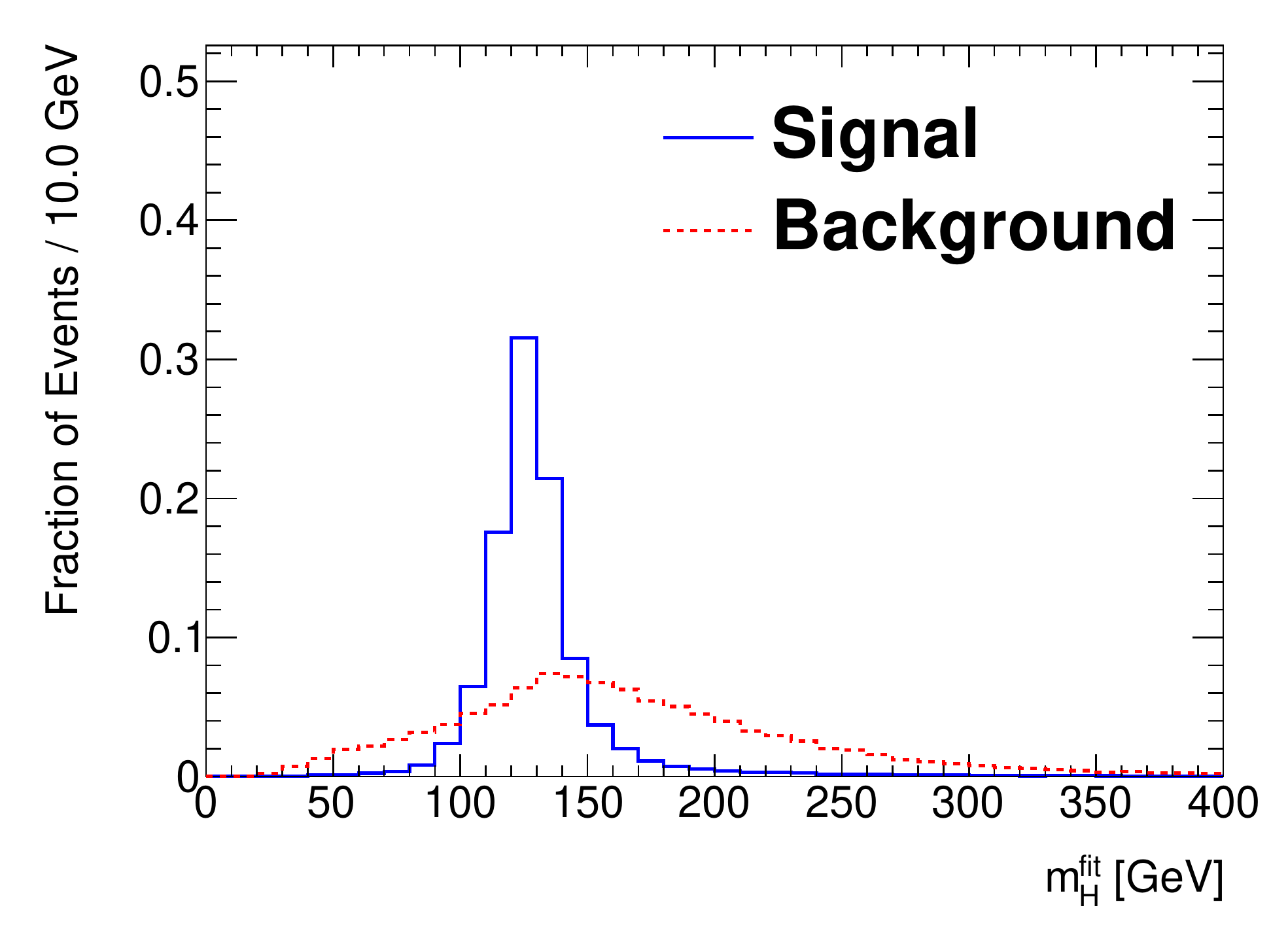}}
\put(-130, 100){\textbf{(b)}}
\subfigure{\includegraphics[width=0.33\textwidth]{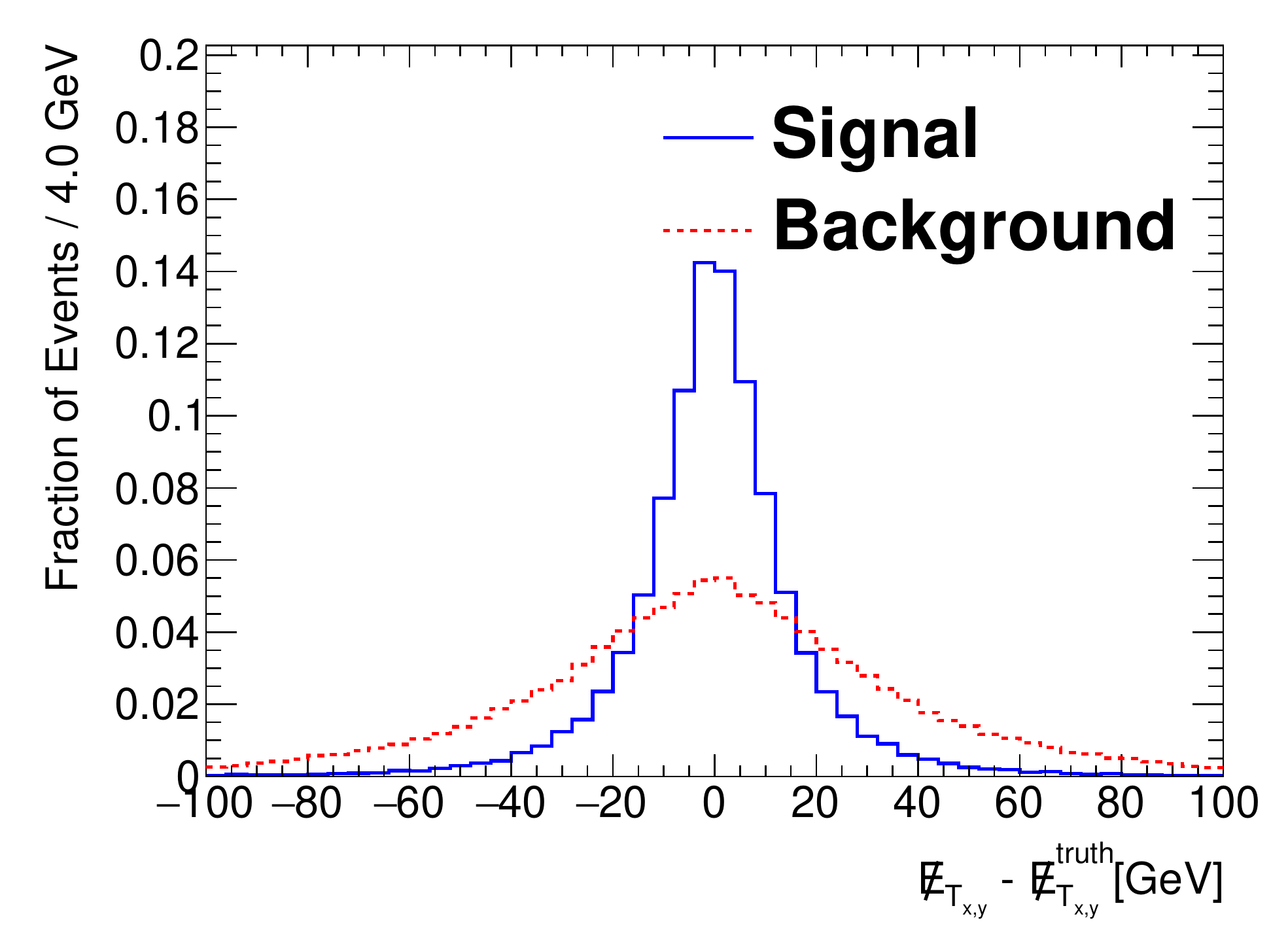}}
\put(-130, 100){\textbf{(d)}}
\subfigure{\includegraphics[width=0.33\textwidth]{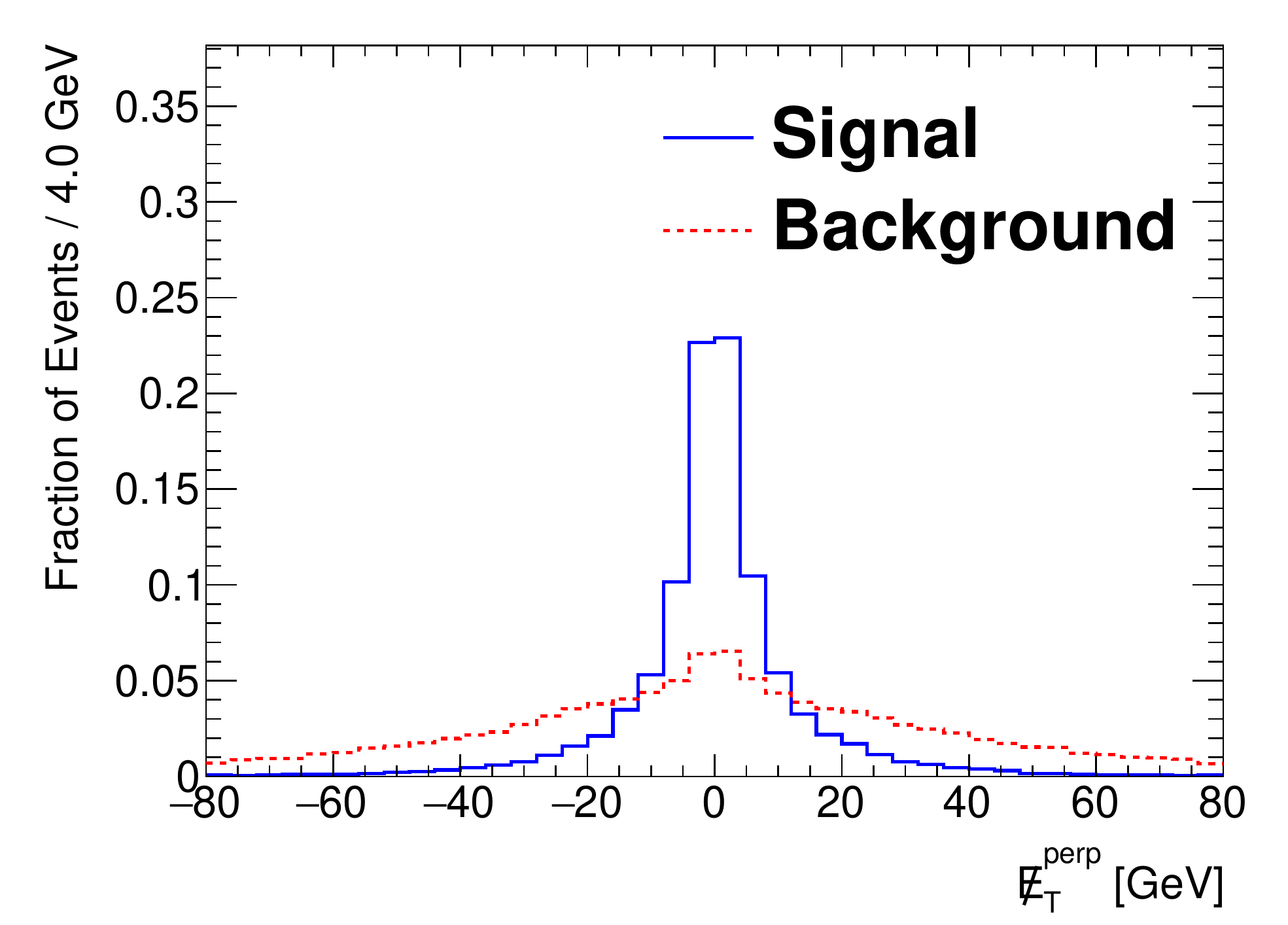}}
\put(-130, 100){\textbf{(f)}}
\caption{\label{fig:mass} (a) and (b) show the distributions of the reconstructed Higgs mass using the collinear approximation and the kinematic fit in this work respectively. (c) and (d) shows the resolutions of the missing transverse energy before and after the kinematic fit respectively. (e) and (f) show the distributions of $\Delta\slashed{E}_{\text{T}}^{\text{proj}}$ and $\slashed{E}_{\text{T}}^{\text{perp}}$ respectively. The blue-solid (red-dashed) histograms represent the signal (background) distributions. }
\end{figure*}
Figure~\ref{fig:mass} shows the performance of the kinematic constraints. In Fig.~\ref{fig:mass}, (a) and (b) show the distributions of the reconstructed Higgs mass using the collinear approximation and in this work respectively. The reconstructed Higgs mass resolution is improved from 16 GeV using the collinear approximation, to 11 GeV after the kinematic fit. Fig.~\ref{fig:mass}(c) and (d) show that the resolution of the $\slashed{E}_{\text{T}}$ improves from 14 GeV before the kinematic fit, to 11 GeV after it for the signal. Whereas for the background, it deteriorates from 14 GeV to 19 GeV. It is also reflected in the distributions of the $\slashed{E}_{\text{T}}$ projections, $\Delta\slashed{E}_{\text{T}}^{\text{proj}}$ and $\slashed{E}_{\text{T}}^{\text{perp}}$. $\Delta\slashed{E}_{\text{T}}^{\text{proj}}$ is defined as the difference between the fitted $\slashed{E}_{\text{T}}$ component projected in the direction of the measured $\slashed{E}_{\text{T}}$ and the measured $\slashed{E}_{\text{T}}$ itself. $\slashed{E}_{\text{T}}^{\text{perp}}$ is defined as the component of the fitted $\slashed{E}_{\text{T}}$ perpendicular to the direction of the measured $\slashed{E}_{\text{T}}$. The distributions of $\Delta\slashed{E}_{\text{T}}^{\text{proj}}$ and $\slashed{E}_{\text{T}}^{\text{perp}}$ are shown in Fig.~\ref{fig:mass}(e) and (f) respectively. 

\begin{figure}[htb]
\centering
\subfigure{\includegraphics[width=0.4\textwidth]{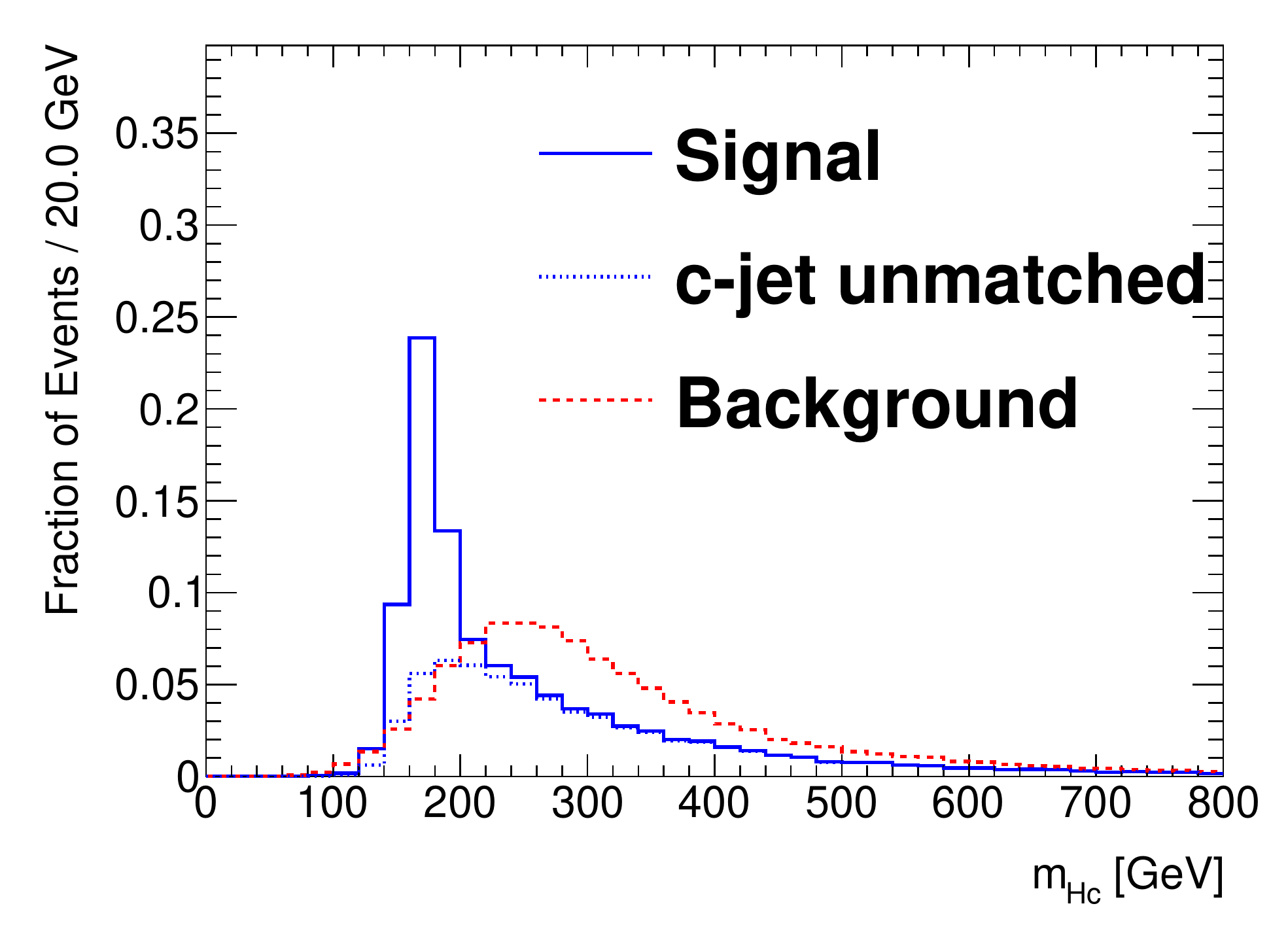}}
\caption{\label{fig:mtop} The distributions of the reconstructed top mass  in the decay $t\to Hc$. The blue-solid (red-dashed) histogram represents the signal (background). The blue-dotted histogram represents the signal with the $c$-jet not matched to the $c$-parton.}
\end{figure}

Finally, Fig.~\ref{fig:mtop} shows the distributions of the reconstructed top mass ($m_{Hc}$) from the decay $t\to Hc$, in the signal and background. We can see that it is a good quantity to distinguish the signal from the background. In Fig.~\ref{fig:mtop} the top mass distribution with the $c$-jet failing to match the $c$-parton in the signal is also shown.  For signal events with $c$-jet matched to the truth, the reconstructed top has a mass resolution of 14 GeV.

\section{Results based on the Multi-Variate Analysis}
\begin{figure*}
\centering
\subfigure{\includegraphics[width=0.33\textwidth]{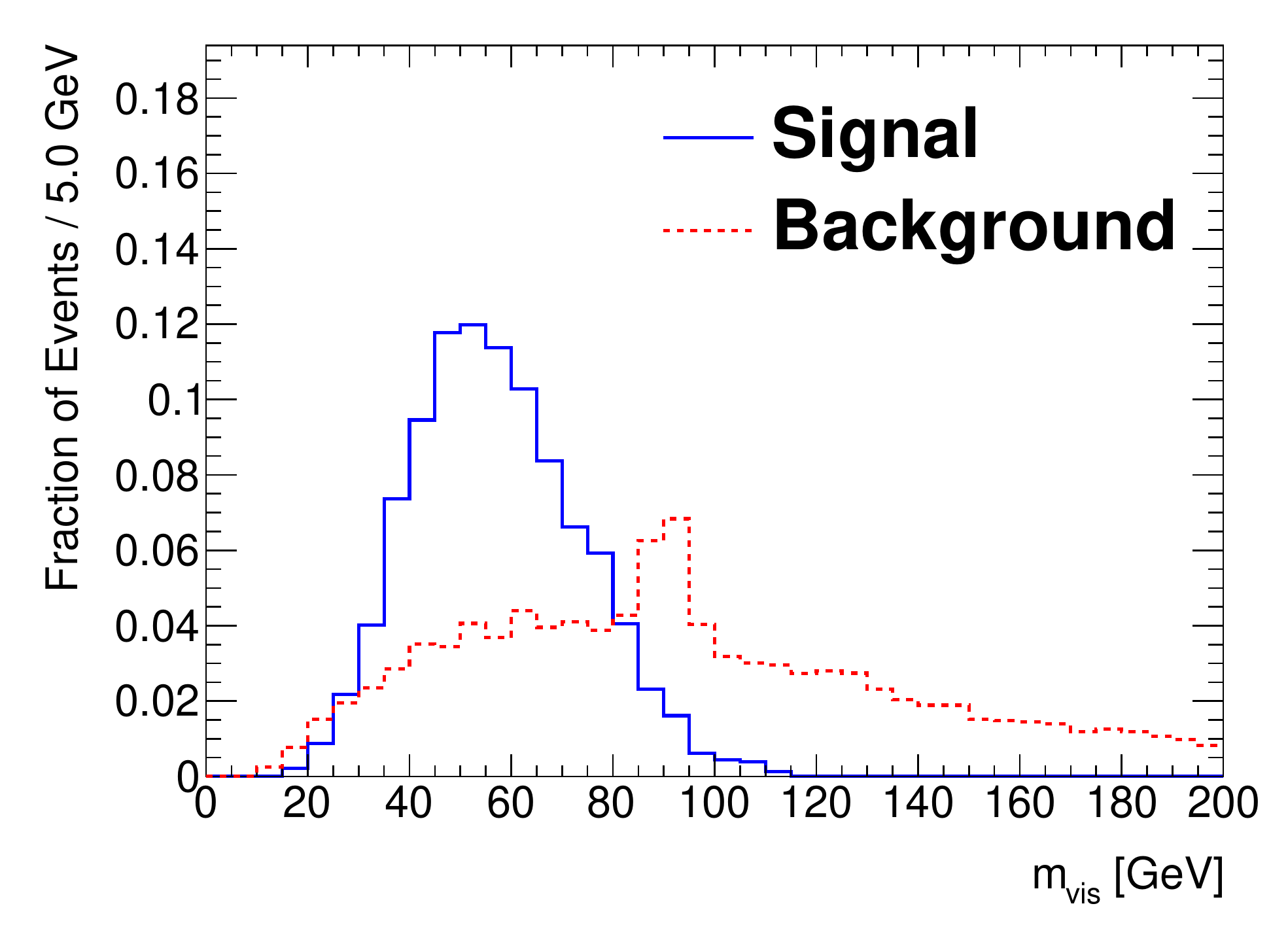}}
\put(-130, 100){\textbf{(a)}}
\subfigure{\includegraphics[width=0.33\textwidth]{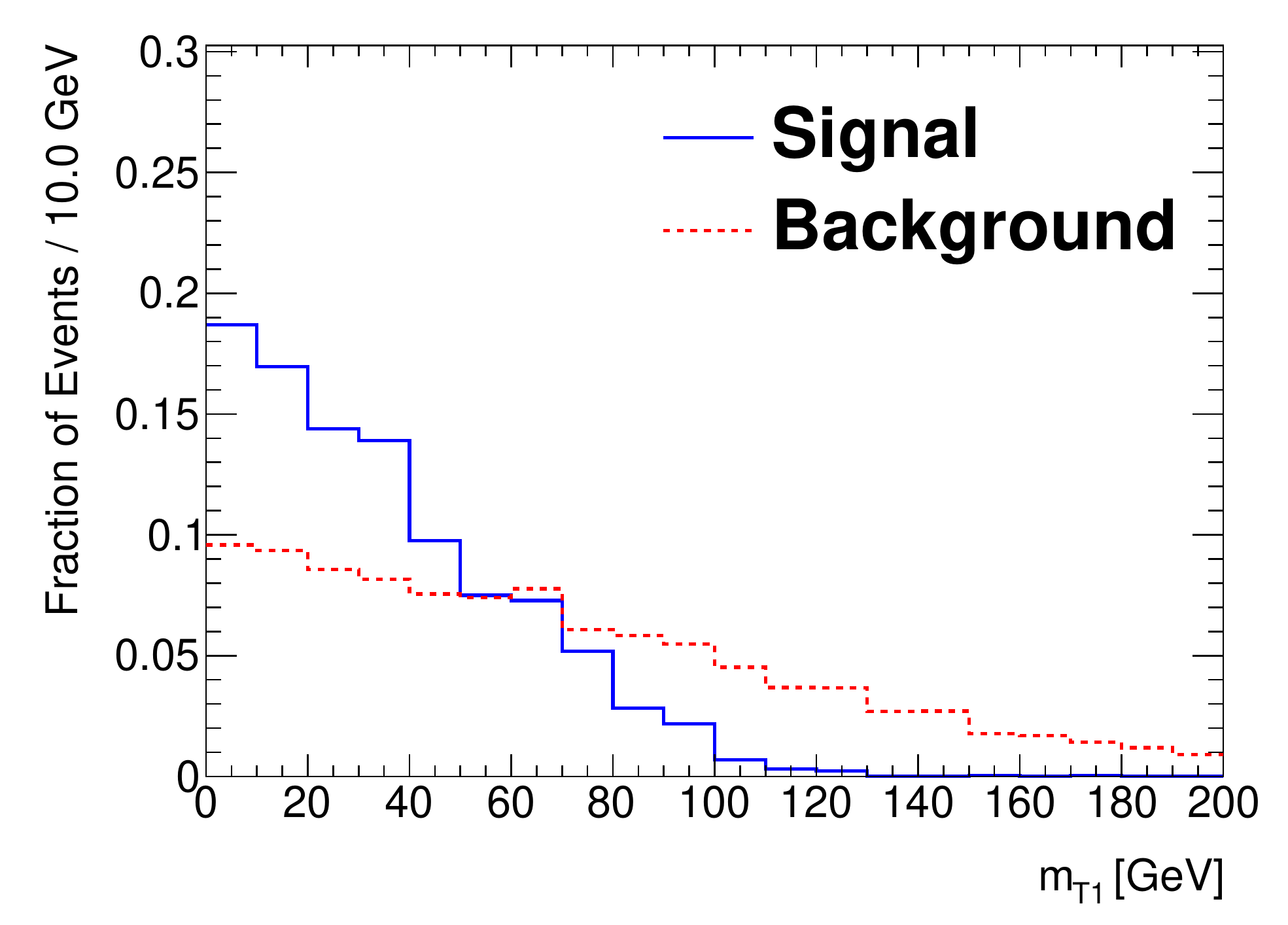}}
\put(-130, 100){\textbf{(b)}}
\subfigure{\includegraphics[width=0.33\textwidth]{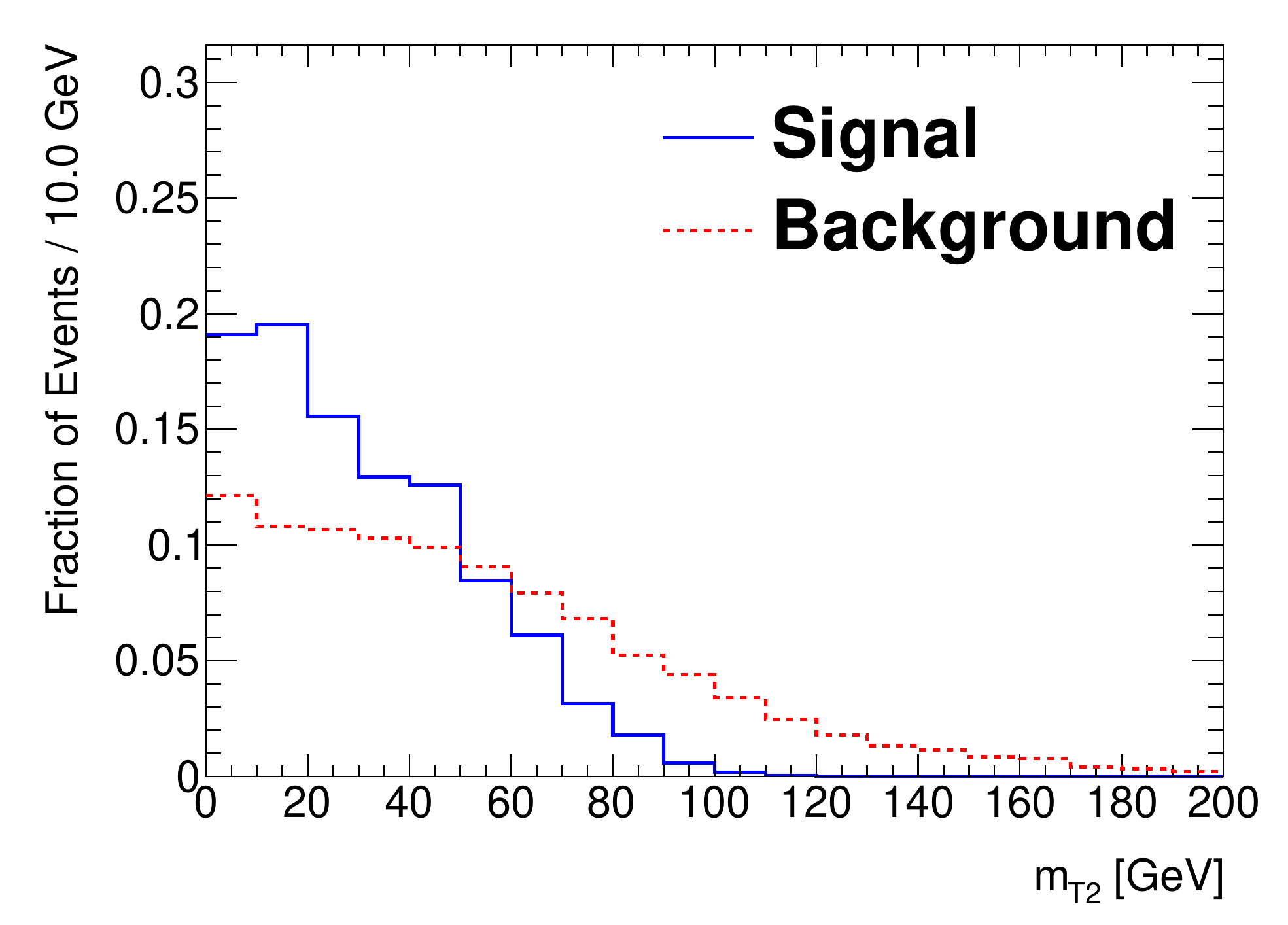}}
\put(-130, 100){\textbf{(c)}}\\
\subfigure{\includegraphics[width=0.33\textwidth]{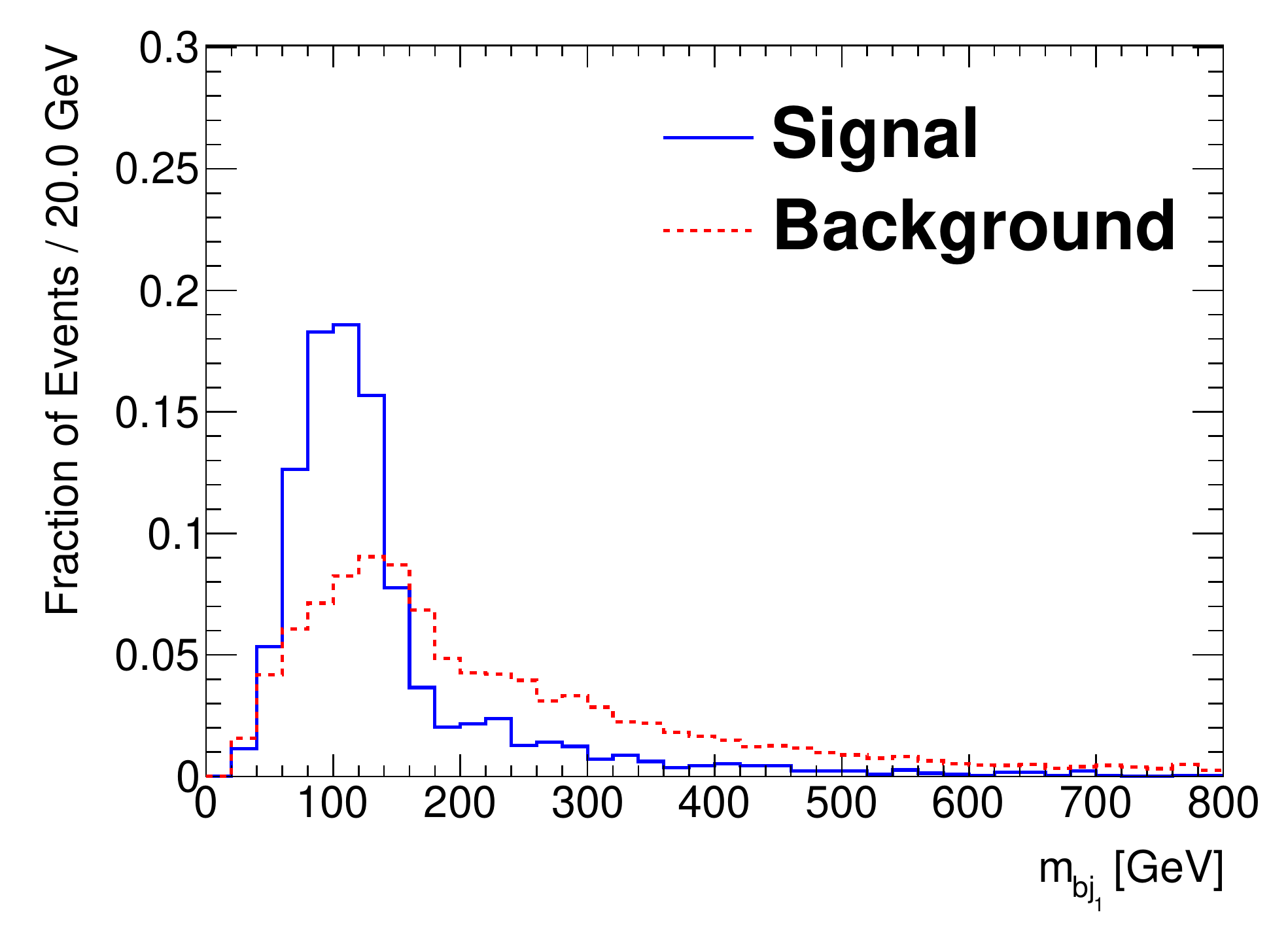}}
\put(-130, 100){\textbf{(d)}}
\subfigure{\includegraphics[width=0.33\textwidth]{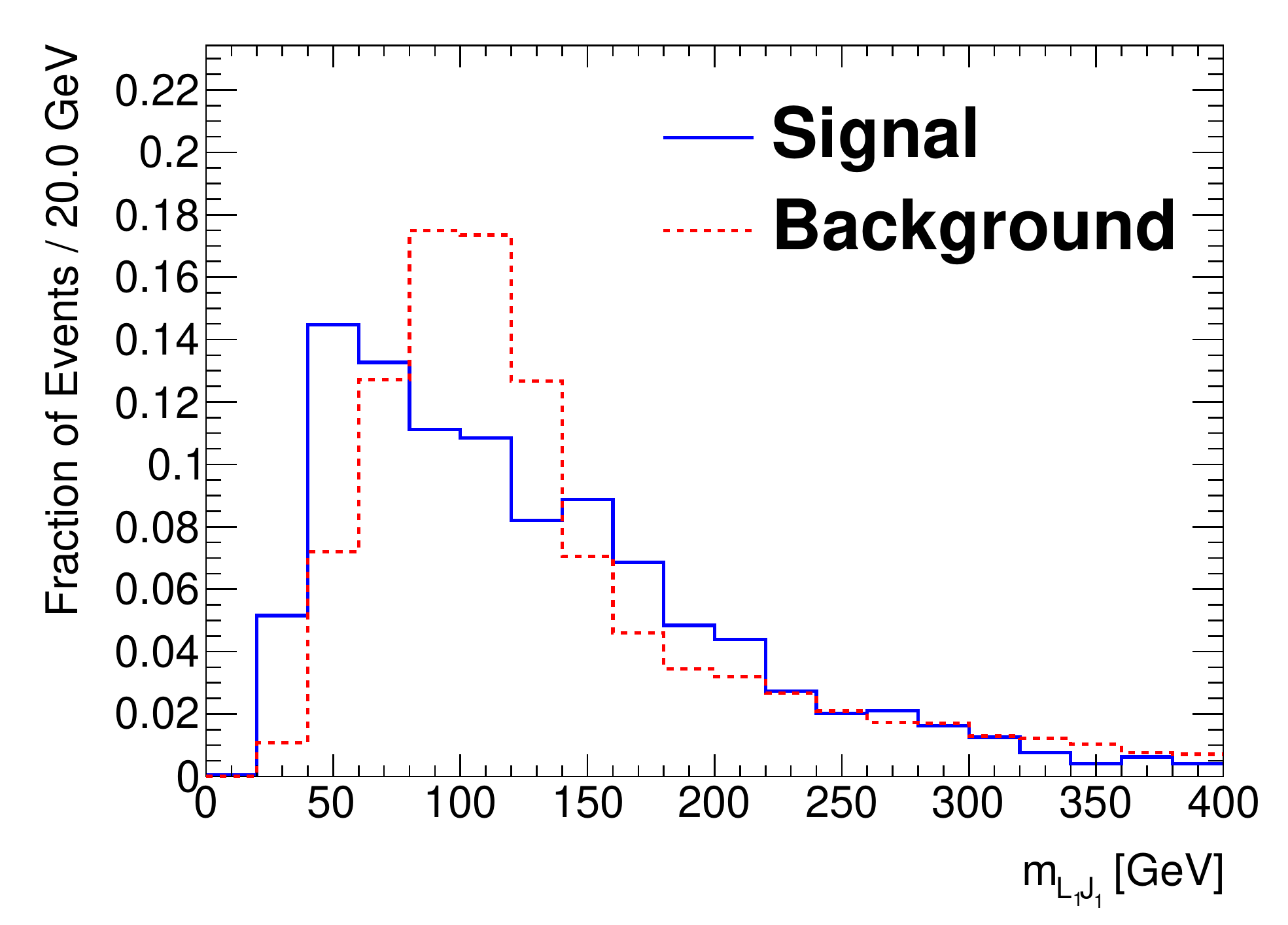}}
\put(-130, 100){\textbf{(e)}}
\subfigure{\includegraphics[width=0.33\textwidth]{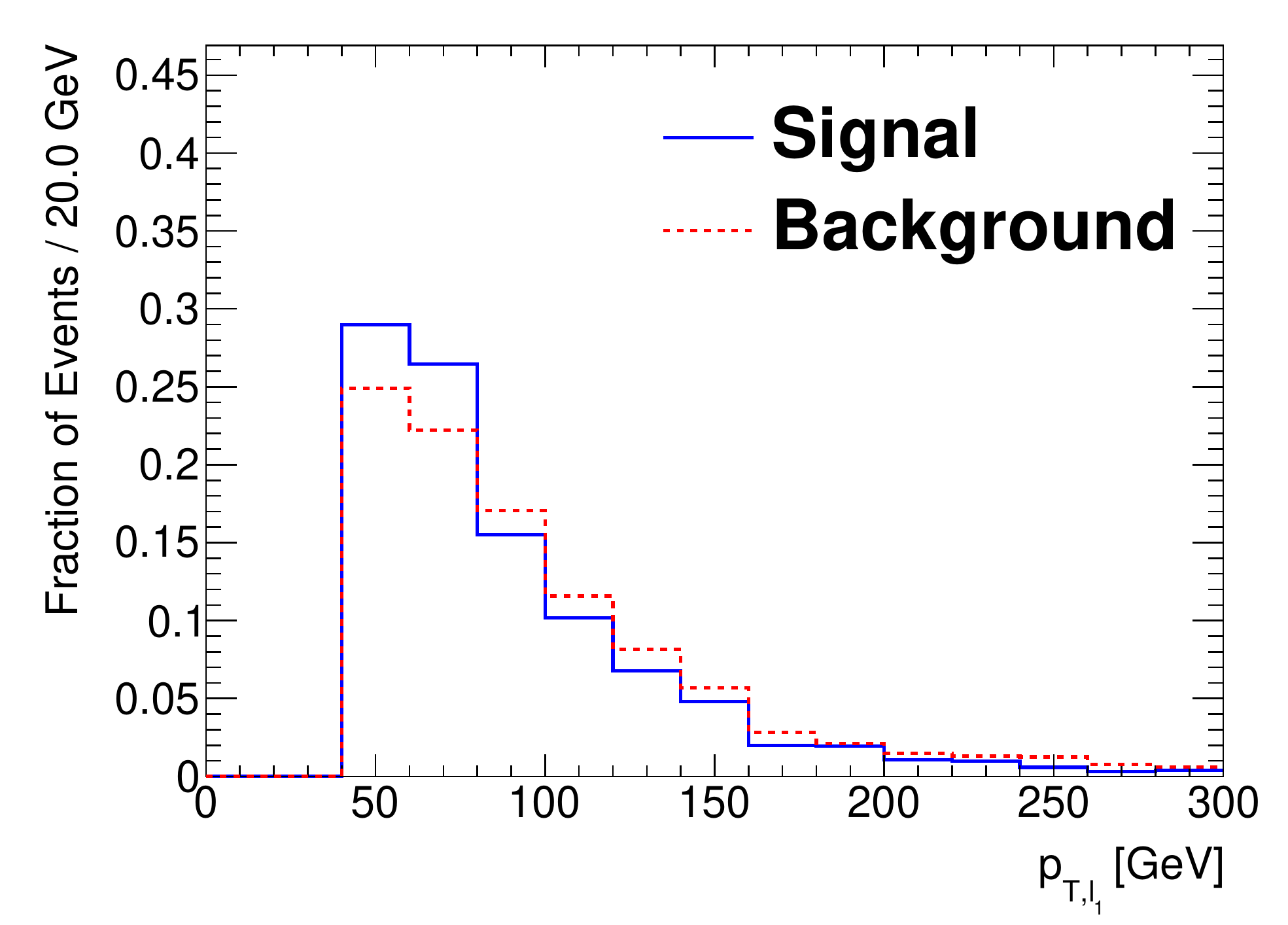}}
\put(-130, 100){\textbf{(f)}}\\
\subfigure{\includegraphics[width=0.33\textwidth]{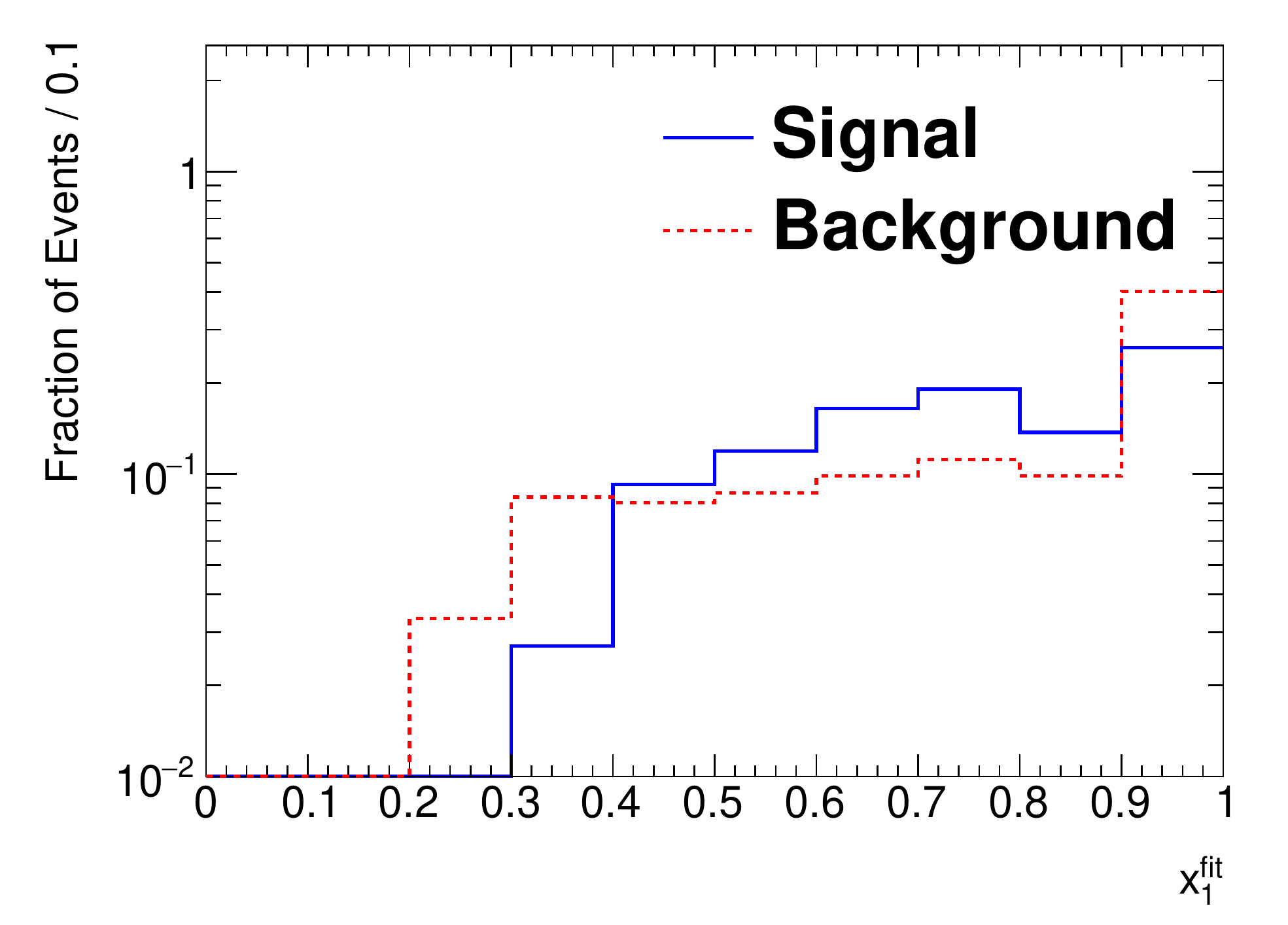}}
\put(-130, 100){\textbf{(g)}}
\subfigure{\includegraphics[width=0.33\textwidth]{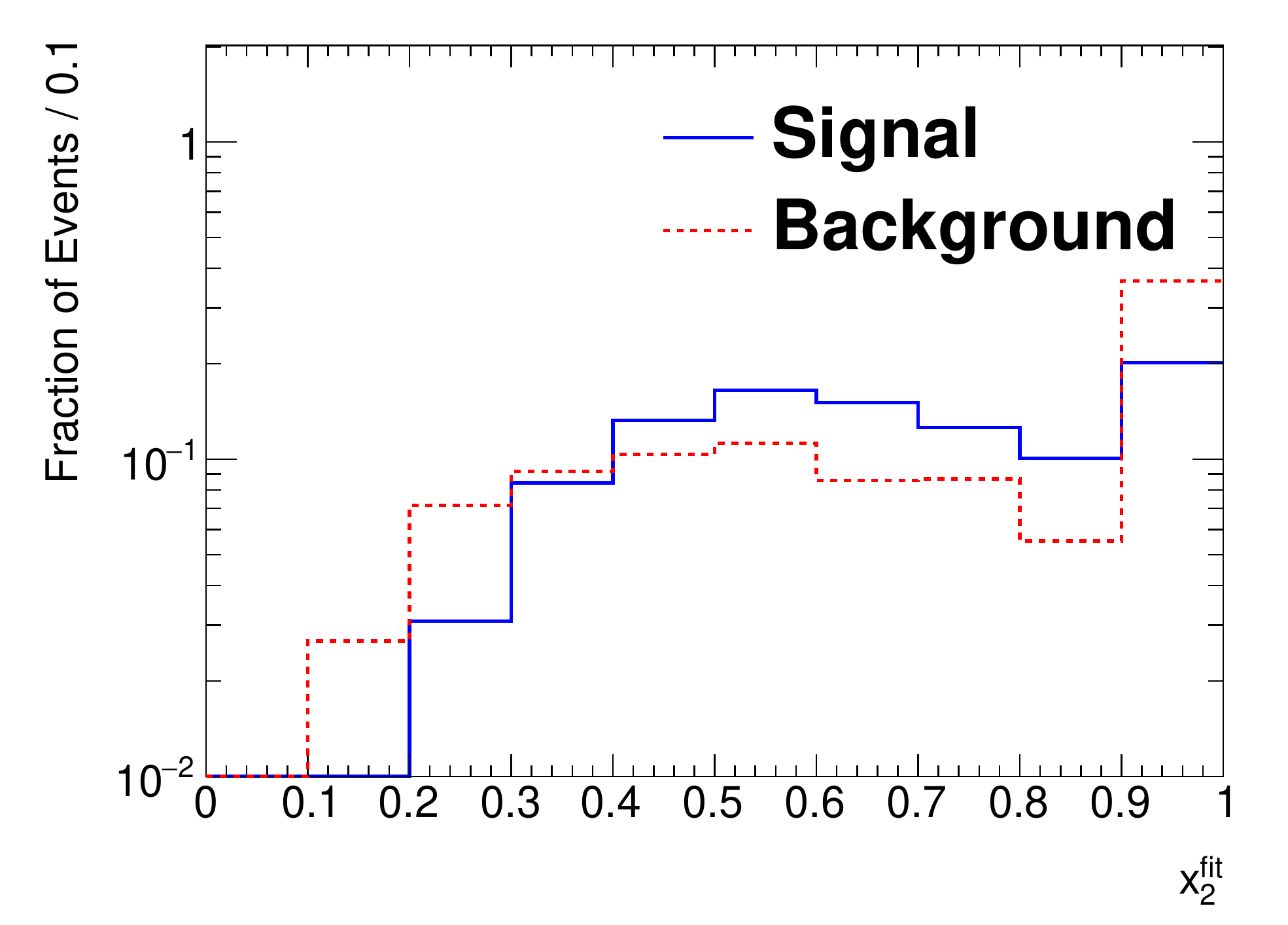}}
\put(-130, 100){\textbf{(h)}}
\subfigure{\includegraphics[width=0.33\textwidth]{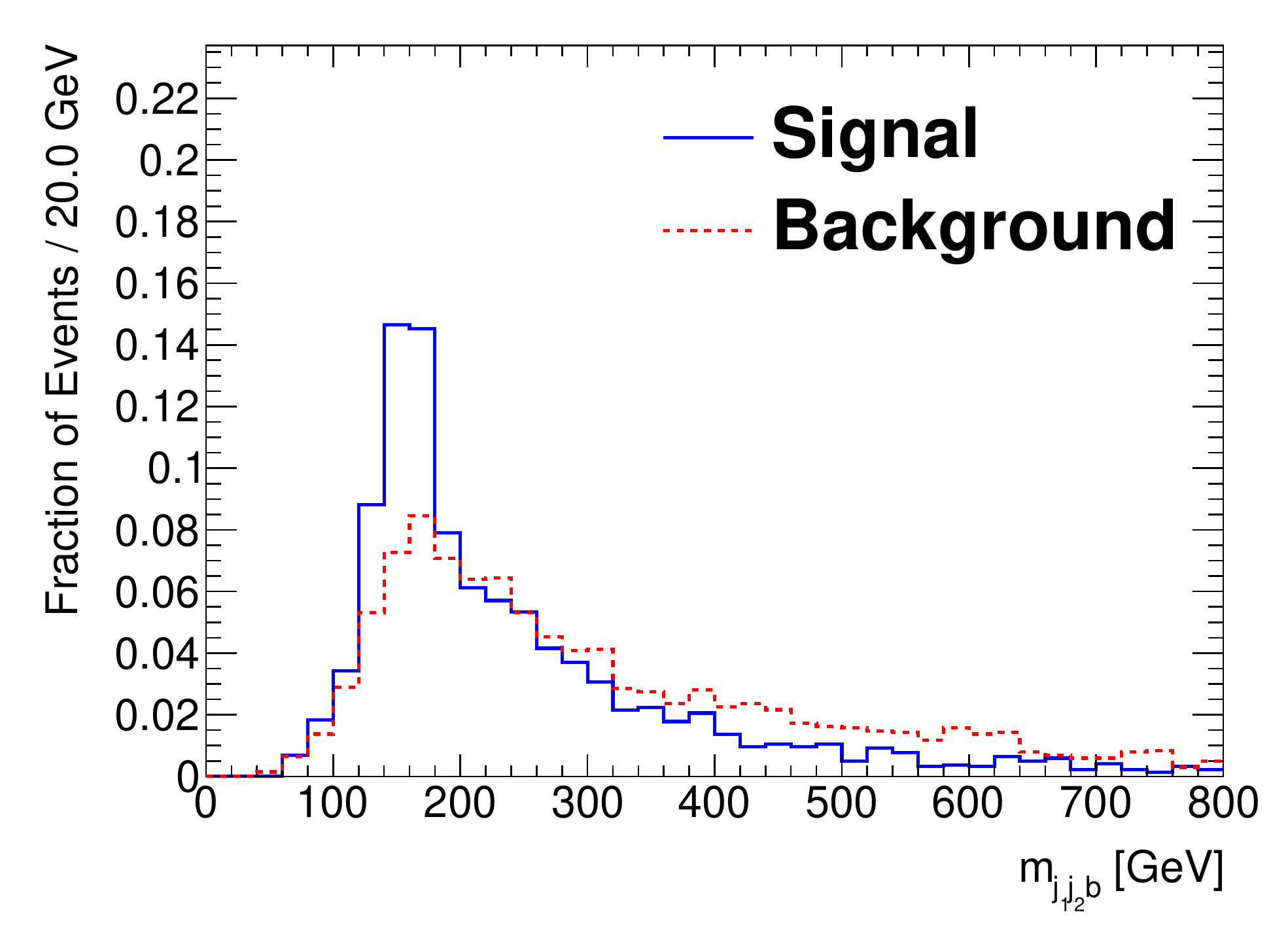}}
\put(-130, 100){\textbf{(i)}}
\caption{\label{fig:mva} The distributions of the BDT input variables: (a) $m_{\text{vis}}$, (b) $m_{\text{T}1}$, (c) $m_{\text{T}2}$, (d) $m_{bj_1}$, (e) $m_{L_1J_1}$, (f) $p_{\text{T},l_1}$, (g) $x_1^{\text{fit}}$, (h) $x_2^{\text{fit}}$ and (i) $m_{j_1j_2b}$. The blue-solid (red-dashed) histograms represent the signal (background) distributions. Here, (a)-(e) are from the $\tau_l\tau_l$ 4-jet category while the others are from the $\tau_h\tau_h$ 4-jet category, for demonstration. }
\end{figure*}

In this section, we investigate the sensitivity of probing $\BR(t\to Hc)$ using one of the Multi-Variate Analysis (MVA) methods, the Gradient Boosted Decision Trees (BDT) method~\cite{BDT,BDT2}.  The BDT output score is in the range between -1 and 1. The most signal-like events have scores near 1 while the most background-like events have scores near -1.
 As shown in Tab.~\ref{tab:tab2}, the signal purity is different for different decay modes ($\tau_l\tau_l$, $\tau_l\tau_h$ and $\tau_h\tau_h$) and for different event topologies (3-jet events and 4-jet events). To maximize the overall sensitivity, the signal region is thus divided into 6 categories, shown in Tab.~\ref{tab:tab2}. In each of them, the Gradient BDT method is used for signal-background separation. A number of variables as the BDT inputs are used to train and test events in each signal region for maximal signal acceptance and background rejection. They are listed in Tab.~\ref{tab:tab1}. Here are the definitions of these variables which are not yet introduced:
 \begin{enumerate}
\item $m_{\text{vis}}$ is the invariant mass of the visible decay products of the tau lepton pair. As shown in Fig.~\ref{fig:mva}(a), the $t\bar{t}$ events have a much wider $m_{\text{vis}}$ distribution than the signal, whereas the $Z$ background events give rise to a small peak around the $Z$ mass. 
\item $m_{\text{T1}}$ ($m_{\text{T2}}$) is the transverse mass calculated from the leading (sub-leading) tau candidate and the $\slashed{E}_{\text{T}}$ for the $\tau_l\tau_l$ and $\tau_h\tau_h$ modes. In the $\tau_l\tau_h$ mode, $m_{\text{T1}}$ ($m_{\text{T2}}$) is the transverse mass from the charged lepton ($\tau$-jet) and the $\slashed{E}_{\text{T}}$. The background distribution is wider than the signal distribution as shown in Fig.~\ref{fig:mva}(b) and~(c).
\item $m_{bj_1}$ is the invariant mass of the $b$-jet and the leading jet presumably from the $W$ decay. As shown in Fig.~\ref{fig:mva}(d), for the signal events, $m_{b j_1}$ is most likely smaller than the top mass while the background events have a wider distribution.
\item $m_{l_1j_1}$ is the invariant mass of the leading tau candidate and the jet which has the smallest angular distance with the tau candidate. For the $t\bar{t}$ events with both tops decaying leptonically, $m_{l_1j_1}$ tends to be smaller than the top mass, as indicated by Fig.~\ref{fig:mva}(e).
\item $p_{\text{T},l_1}$ ($p_{\text{T},l_2}$) is the transverse momentum of the leading (sub-leading) tau candidate ($e/\mu/\tau$-jet). A example of the $p_{\text{T},l_1}$ distribution is shown in Fig.~\ref{fig:mva}(f). 
\item $x_{1,2}^{\text{fit}}$ is the momentum fraction carried by the visible decay products of the tau calculated using the best-fit 4-momentum of the neutrino(s). For the $\tau_h$ decay mode, the visible decay products carry most of the tau energy since there is only a single neutrino in the final state, which is evident in the excess around 1 in Fig.~\ref{fig:mva}(g) and~(h). 
\item $m_{j_1j_2b}$ is the invariant mass of the $b$-jet and the two jets from the $W$ decay and is the top mass for the decay $t\to Wb \to j_1j_2b$. In the signal events, the top mass is visible as in Fig.~\ref{fig:mva}(i).
\end{enumerate}
\begin{table}
\caption{The BDT input variables (checkmarked) used in each signal region. }
\begin{ruledtabular}
\begin{tabular}{ccccccc}
 & \multicolumn{2}{c}{$\tau_l\tau_l$} & \multicolumn{2}{c}{$\tau_l\tau_h$} & \multicolumn{2}{c}{$\tau_h\tau_h$} \\ \hline
 & 3-jet & 4-jet & 3-jet & 4-jet & 3-jet & 4-jet \\ \hline
$m_{H}^{\text{fit}}$                          	& \checkmark  & \checkmark  & \checkmark  & \checkmark  & \checkmark  & \checkmark \\
$\Delta \slashed{E}_{\text{T}}^{\text{proj}}$ 	& \checkmark  & \checkmark  & \checkmark  & \checkmark  & \checkmark  & \checkmark \\
$\slashed{E}_{\text{T}}^{\text{perp}}$          	& \checkmark  & \checkmark  & \checkmark  & \checkmark  & \checkmark  & \checkmark \\
$m_{\text{vis}}$                                	& \checkmark  & \checkmark  & \checkmark  & \checkmark  &  &  \\
$m_{Hc}$                                	& \checkmark  & \checkmark  & \checkmark  & \checkmark  & \checkmark  & \checkmark \\
$m_{\text{T1}}$                                	& \checkmark  & \checkmark  & \checkmark  & \checkmark  & \checkmark  & \\
$m_{\text{T2}}$                                	& \checkmark  & \checkmark  &   		    &   		 &   & \\
$C_{\text{miss}}$                              	& \checkmark  & \checkmark  & \checkmark  & \checkmark  & \checkmark  & \checkmark  \\
$m_{b j_1}$                   		& \checkmark  & \checkmark  & \checkmark  & \checkmark  & \checkmark  & \checkmark \\
$p_{\text{T},l_1}$                      		&  			&   			&                     &                      & \checkmark  & \checkmark \\
$p_{\text{T},l_2}$                      		&  			&   			& \checkmark  & \checkmark  & \checkmark  & \checkmark \\
$m_{l_1j_1}$      			& \checkmark   & \checkmark  &   			&   			&   			&   \\
$x_1^{\text{fit}}$					&			&			&			&		&\checkmark			&\checkmark \\
$x_2^{\text{fit}}$					&			&			&\checkmark 	&\checkmark	&\checkmark			&\checkmark \\
$m_{j_1 j_2 b}$             		& 			& \checkmark  &   			& \checkmark  &  & \checkmark \\
\end{tabular}	
\label{tab:tab1}
\end{ruledtabular}
\end{table}

\begin{figure}
\centering
\subfigure{\includegraphics[width=0.4\textwidth]{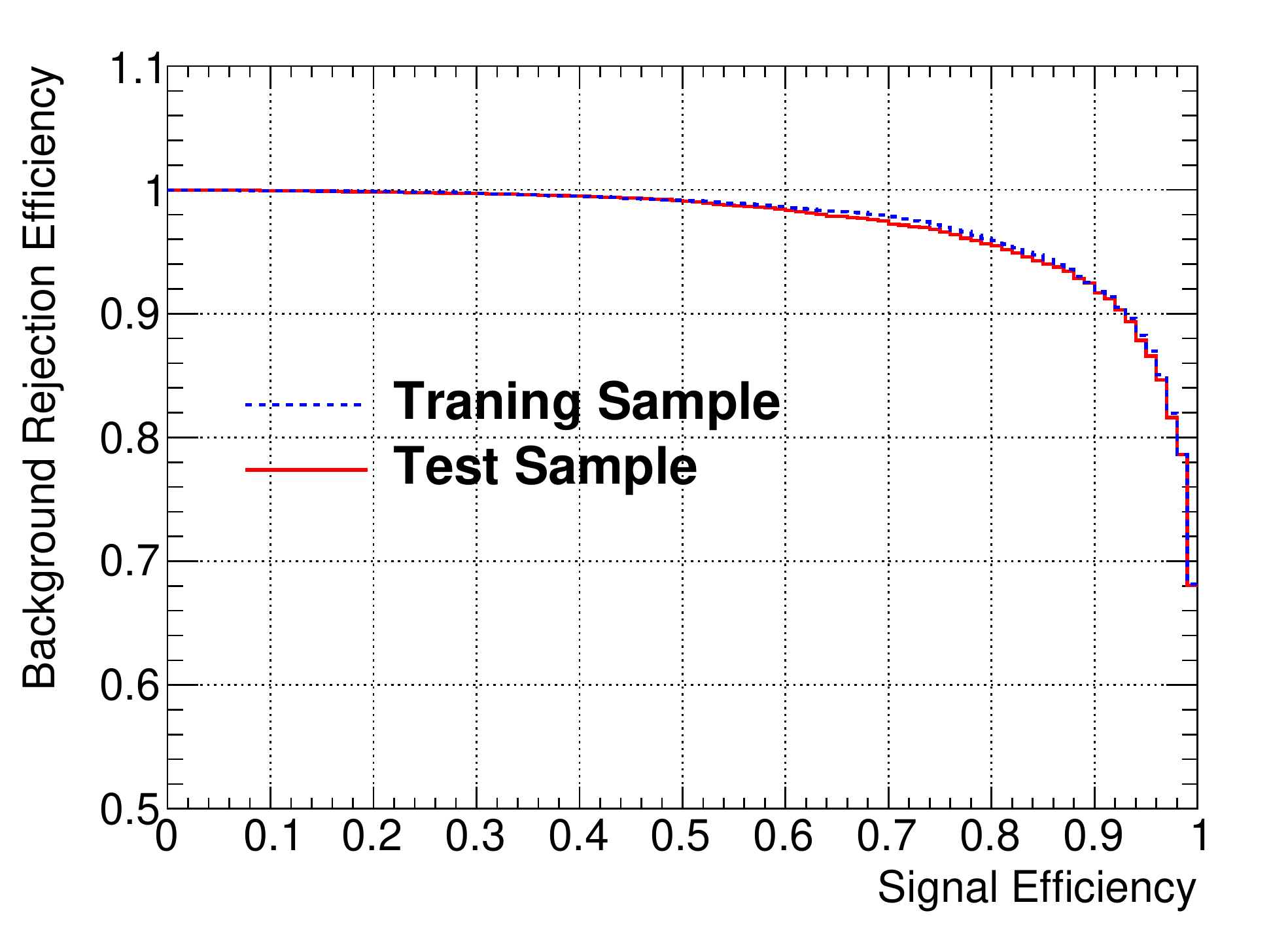}}
\caption{\label{fig:roc} The signal acceptance versus background rejection efficiency curves from the training (blue dashed) and test (red solid) samples in the $\tau_l \tau_l$ 4-jet category. }
\end{figure}

\begin{figure*}
\centering
\subfigure{\includegraphics[width=0.4\textwidth]{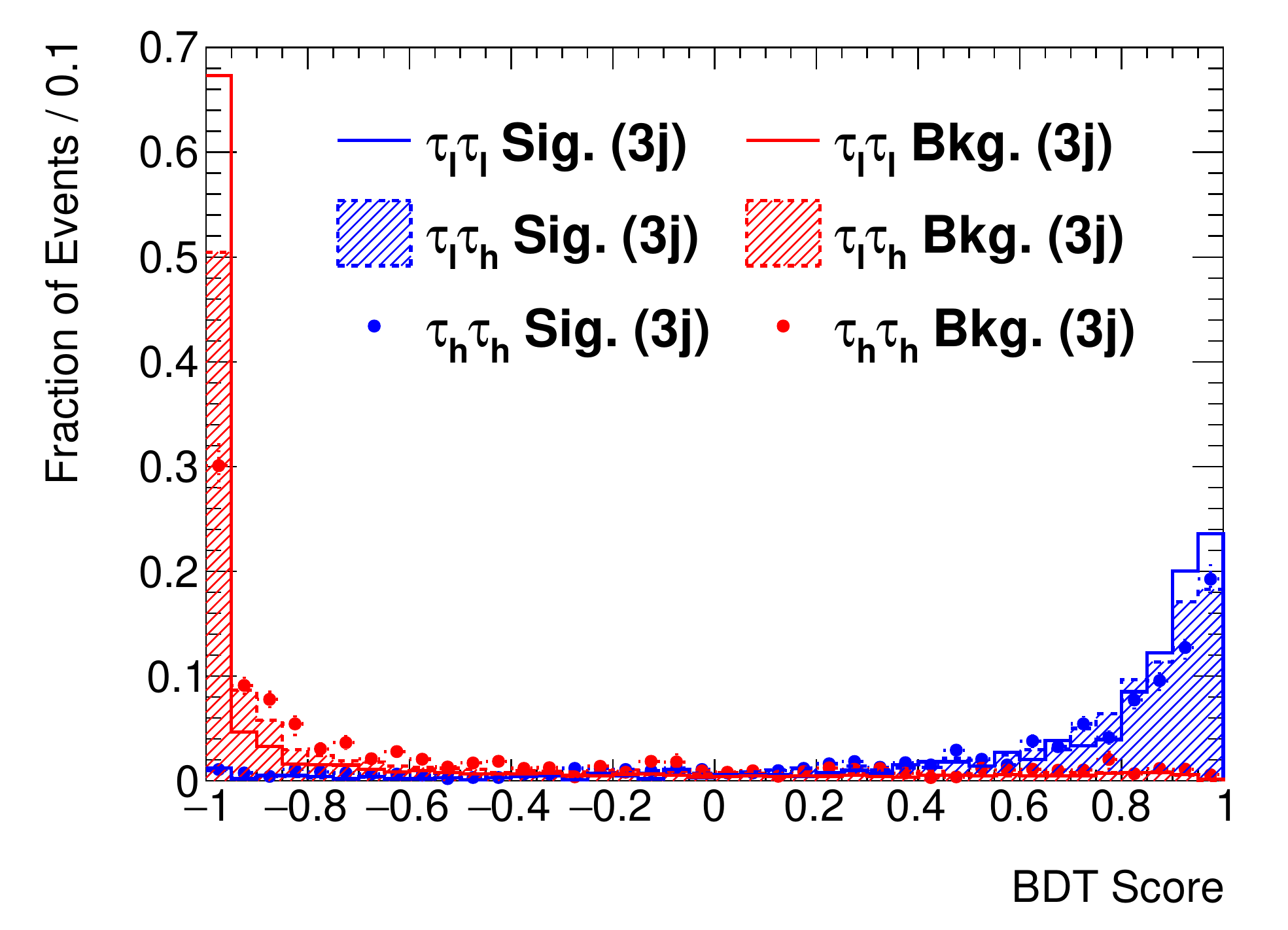}}
\put(-165, 125){\textbf{(a)}}
\subfigure{\includegraphics[width=0.4\textwidth]{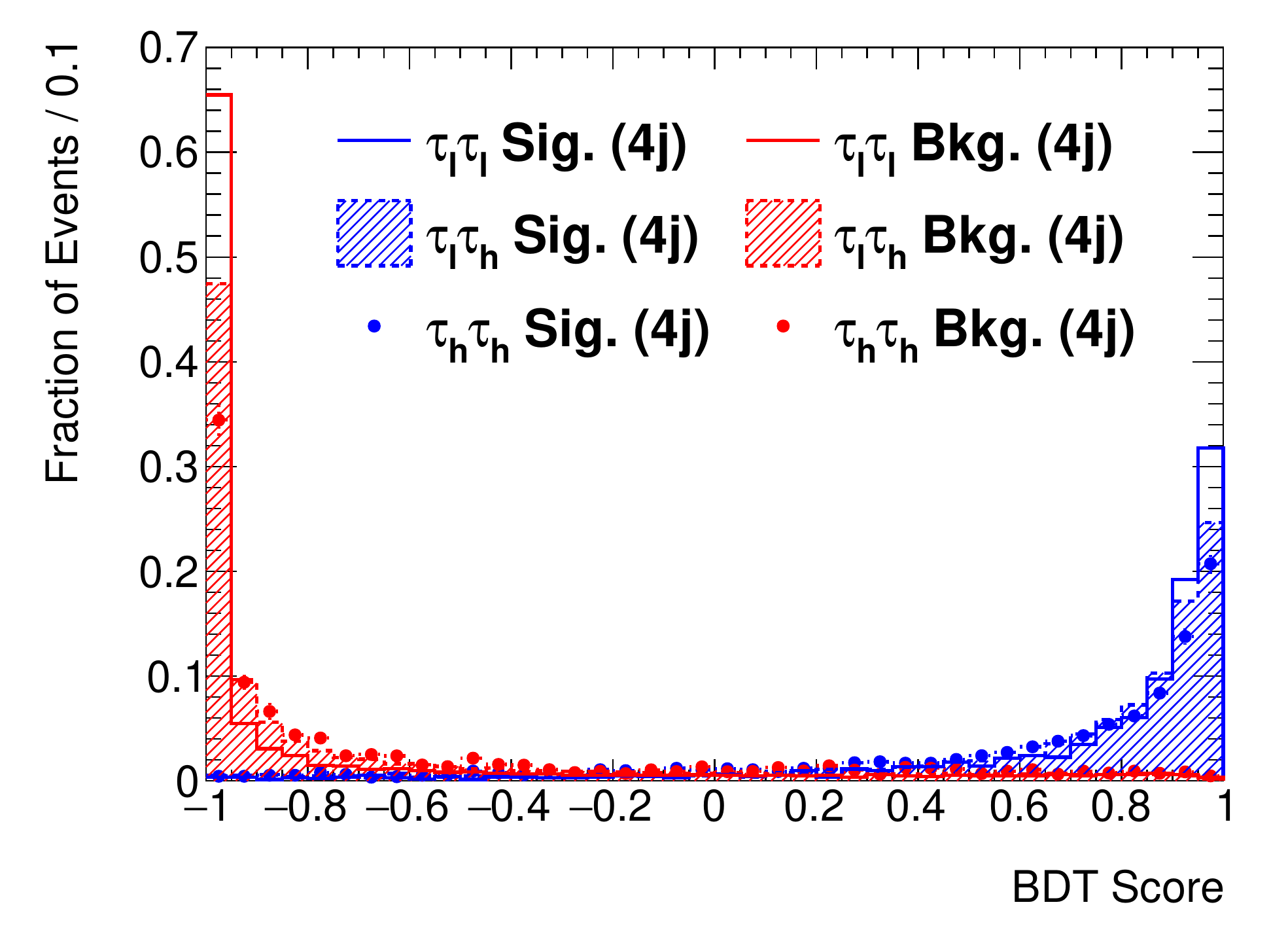}}
\put(-165, 125){\textbf{(b)}}\\
\caption{\label{fig:BDT} The distributions of the BDT score in the 3-jet (a) and 4-jet (b) events. The blue (red) histograms represent the signal (background) distributions. The solid histograms, hatched histograms and dots represent the decay modes $\tau_l\tau_l$, $\tau_l\tau_h$ and $\tau_h\tau_h$, respectively. }
\end{figure*}

The signal and background samples are divided into two halves, with one for BDT training and the other for testing. The normalized BDT score distributions are shown in Fig.~\ref{fig:BDT}(a)-(b). The signal acceptance versus background rejection efficiency curve based on the BDT distributions in the $\tau_l \tau_l$ 4-jet category is shown in Fig.~\ref{fig:roc} as a demonstration. The curves from both the training and test samples are shown in the same figure. The BDT parameters are set such that overtraining is not serious, and at the same time the BDT performance is not much compromised. The final results are based on the BDT distributions from the test samples. To highlight the importance of the variables $\Delta \slashed{E}_{\text{T}}^{\text{proj}}$, $\slashed{E}_{\text{T}}^{\text{perp}}$ and the Higgs mass constraint in Eq.~\ref{eq:eq2} introduced in this analysis, the BDT performance without using them is also studied. It is found that the background rate will increase by about 12\% (25\%) with a signal efficiency of $90\%$ in the $\tau_l\tau_h$ 4-jet category without the Higgs mass constraint (without the Higgs mass constraint, $\Delta \slashed{E}_{\text{T}}^{\text{proj}}$ and $\slashed{E}_{\text{T}}^{\text{perp}}$). 

\begin{figure}
\centering
\includegraphics[width=0.4\textwidth]{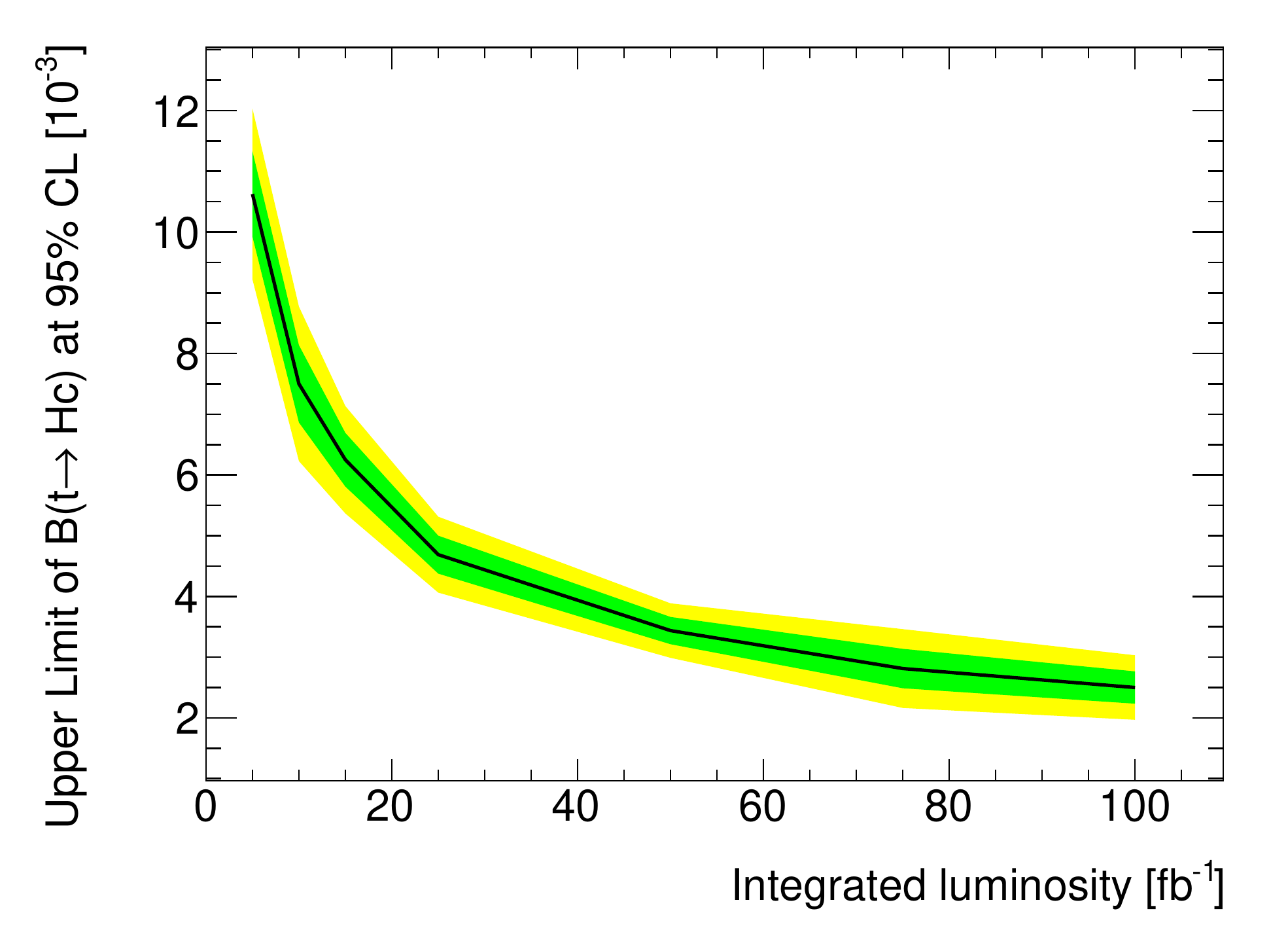}
\caption{ The expected upper limit on $\BR(t\to Hc)$ at 95\% CL level as a function of the integrated luminosity at 13 TeV collision energy.  The green and yellow bands represent the $1\sigma$ and $2\sigma$ systematic uncertainties.}
\label{fig:sens}
\end{figure}

A simultaneous fit is carried out in all six signal regions with the six BDT score discriminants. Since the signal and $t\bar{t}$ background come from the same production process, the signal yield in each region, $n_{\text{sig}}$, can be expressed as a function of the branching ratio, $\BR(t\to Hc)$, and the number of $t\bar{t}$ events, $n_{t\bar{t}}$.
\begin{equation}
\frac{n_{\text{sig}}}{n_{t\bar{t}}} = 2 \BR(H\to\tau\tau) \BR(W\to jj) \frac{\varepsilon_{\text{sig}}}{\varepsilon_{t\bar{t}}} \frac{\BR(t\to Hc)}{1-\BR(t\to Hc)} ,
\label{eq:eq7}
\end{equation}
where $\varepsilon_{\text{sig}}$ and $\varepsilon_{t\bar{t}}$ are the selection efficiencies of the signal and the $t\bar{t}$ background. By taking the relative ratio of the signal to the $t\bar{t}$ background, many systematics, such as the luminosity uncertainty, $t\bar{t}$ production cross section uncertainty, parton-density-function (PDF) uncertainty, and the factorization/renormalization scale uncertainty, can be cancelled or reduced. One of the main systematics is the fraction of fake $\tau$ events, which affects the ratio ${\varepsilon_{\text{sig}}}/{\varepsilon_{t\bar{t}}}$, and can only be studied with dedicated control samples from data. The $Z/\gamma^{\star}\to ll$ background strongly depends on the $b$-tagging efficiency and the mis-tag rate (a light jet tagged as a $b$-jet). It is fixed in the fit, since its fraction is small and its normalization can be determined from the $Z\to ee/\mu\mu$ mass peak. The main systematics affecting the results are the $\tau$ energy scale (TES, 3\%), the jet energy scale (JES, 3\%), and the $H\to \tau\tau$ branching ratio (5.7\%). They contribute to both the BDT shape and event normalization systematics.
To derive the upper limit on the branching ratio of $\BR(t\to Hc)$, the profile likelihood ratio, $q_\mu=-2\ln (L(\mu,\hat{\hat{\theta}}_\mu)/L(\hat{\mu},\hat{\theta}) )$~\cite{q1,q2}, is used as the test statistic, where $\mu$ is just $\BR(t\to Hc)$, and $\theta$ represents the nuisance parameters, namely, the background yields, TES, JES and $\BR(H\to\tau\tau)$. The $\hat{\mu}$ and $\hat{\theta}$ are the parameter values that maximize the likelihood, and $\hat{\hat{\theta}}_\mu$ is the value that maximizes the likelihood for a given value of $\mu$ being scanned. Figure~\ref{fig:sens} shows the expected upper limit at 95\% confidence level (CL) of $\BR(t\to Hc)$, as a function of the integrated luminosity.  The numerical upper limits  are also given in Tab.~\ref{tab:result}. The analysis is able to probe $\BR(t\to Hc)$ down to 0.25\% with a data set of 100~fb$^{-1}$ at $\sqrt{s}=13$ TeV. 

\begin{table}[H]
\caption{The expected upper limits (U.L.) of $\BR(t\to Hc)$ at 95\% Confidence Level for different integrated luminosities.}
\begin{ruledtabular}
\begin{tabular}{llllllll}
Lumi. (fb$^{-1}$) & 5 & 10 & 15 & 25 & 50 & 75 & 100 \\
\hline
U.L. ($10^{-3}$) & 10.6 & 7.5 & 6.3 & 4.7 & 3.4 & 2.8 & 2.5 \\
\end{tabular}
\label{tab:result}
\end{ruledtabular}
\end{table}

To estimate the degree of improvement of our method, we try to scale the CMS/ATLAS experimental results of $t\to Hc$ at 7/8~TeV to their expected values with a luminosity of 100~fb$^{-1}$ at 13~TeV, based on the fact that the upper limit ($\BR^{\text{up}}$) scales with the luminosity and cross section as $\BR^{\text{up}} \propto (\sqrt{\sigma \mathcal{L}})^{-1}$ :
\begin{equation}
\frac{\BR^{\text{up}}(13\text{TeV}, 100\text{fb}^{-1})}{\BR^{\text{up}}(8\text{TeV}, \mathcal{L})} = \sqrt{\frac{\sigma_{t\bar{t}}(8\text{TeV})\times \mathcal{L}}{\sigma_{t\bar{t}}(13\text{TeV})\times 100\text{fb}^{-1}}} .
\label{eq:lumi}
\end{equation}
Table~\ref{tab:lumi} lists the upper limits of $t\to Hc$ in different Higgs decays modes, and their expected values with $100~\text{fb}^{-1}$ at 13 TeV extrapolated from Eq.~\ref{eq:lumi}. The upper limit obtained in this work is better than the CMS/ATLAS results involving $\tau$'s in the final state.

\begin{table}[htbp]
\caption{\label{tab:improve} Comparison of the measured upper limits of $\BR(t\to Hc)$. Note that the ATLAS channel-specific results are read from figures in the corresponding references, thus are only approximate. }
\begin{ruledtabular}
\begin{tabular}{lllc}
& \multirow{2}{*}{Higgs decay} & $\BR^{\text{up}}$ & Scaled $\BR^{\text{up}}$\\ 
 & & 7/8~TeV & 13 TeV/100~fb$^{-1}$ \\
\hline
\multirow{5}{*}{CMS~\cite{fcnc_cms} }  & $h\to WW^*$ & 1.58 \% & 0.39\%\\
 & $h\to\tau\tau$ & 7.01\% & 1.71\%\\
 &$h\to ZZ^*$ & 5.31\% & 1.30\%\\
&$h\to\gamma\gamma$ &  0.69\% & 0.17\%\\
& Combined & 0.56\% & 0.14\% \\
\hline
\multirow{4}{*}{ATLAS~\cite{fcnc_atlas} }  & $h\to WW^*/\tau\tau$ & $\sim 0.8\%$ & 0.21\%\\
& $h\to\gamma\gamma$ & $\sim 0.79\%$ & 0.21\%\\ 
 & $h\to b\bar{b}$ &  $\sim 0.56\%$ & 0.15\%\\
& Combined & 0.46\% & 0.12\%\\
\hline
This work & $h\to\tau\tau$ & & 0.25\% \\ 
\end{tabular}
\end{ruledtabular}
\label{tab:lumi}
\end{table}

\begin{figure*}
\centering
\subfigure{\includegraphics[width=0.4\textwidth]{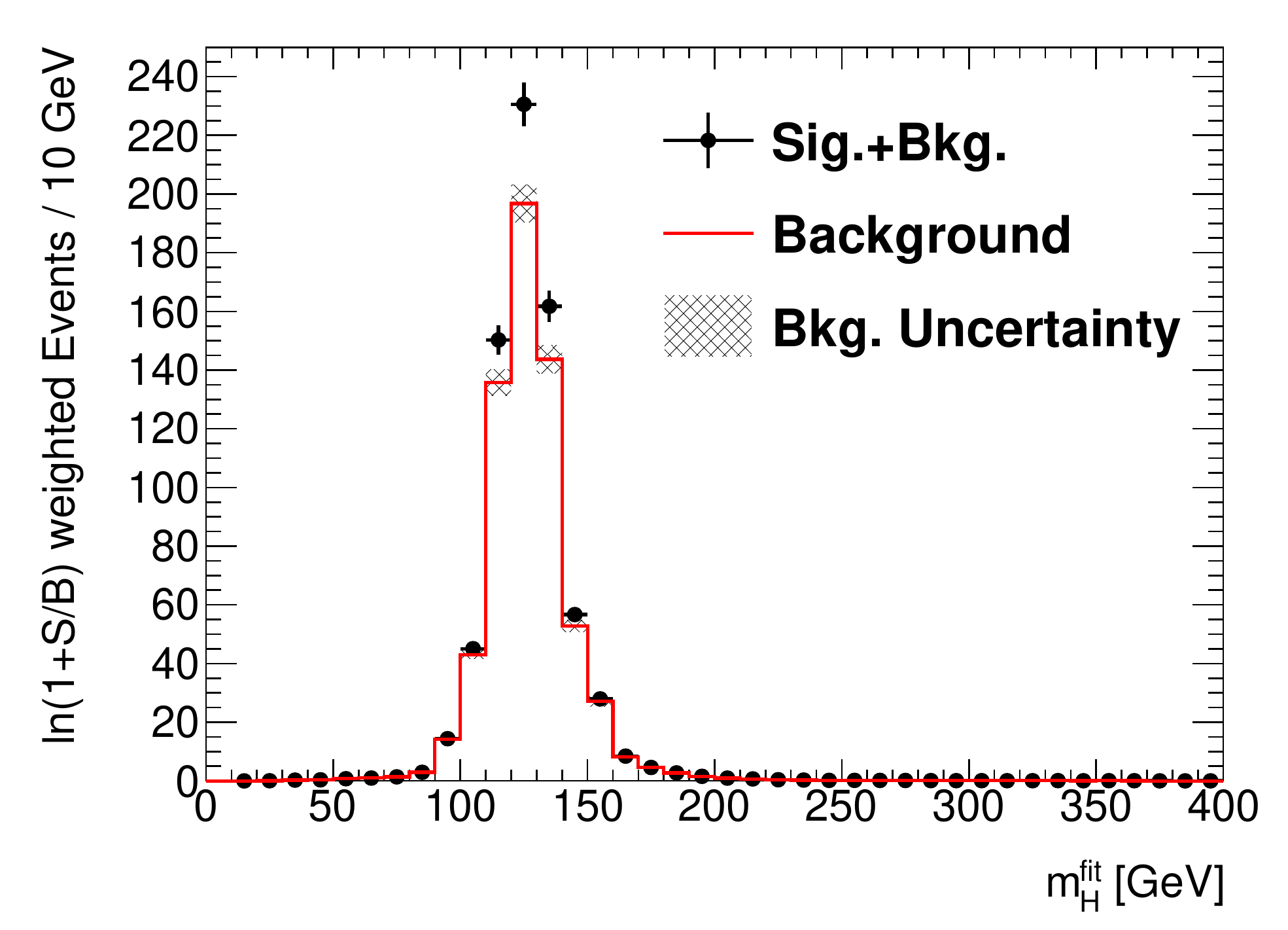}}
\put(-160, 120){\textbf{(a)}}
\subfigure{\includegraphics[width=0.4\textwidth]{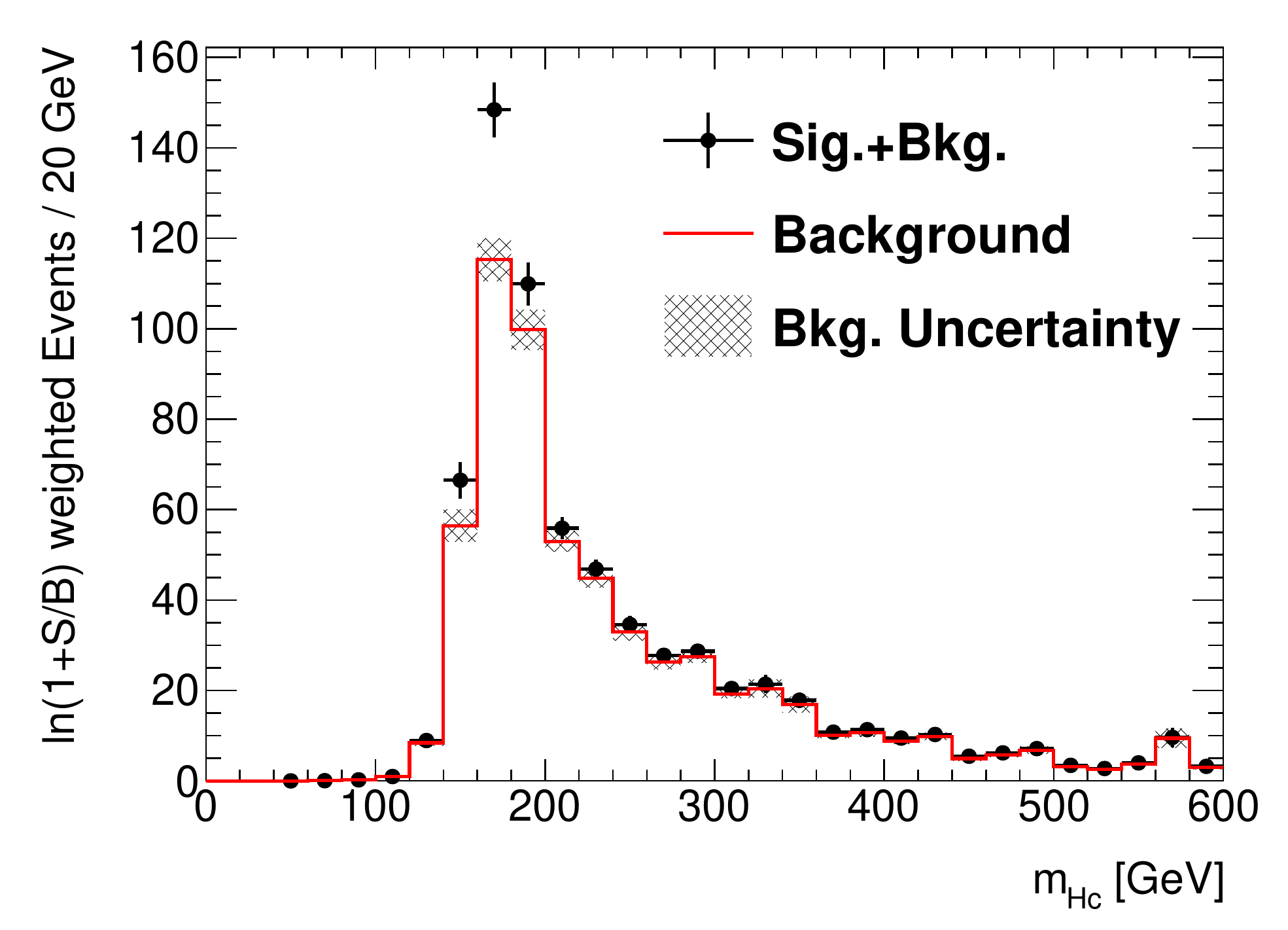}}
\put(-160, 120){\textbf{(b)}}
\caption{ The expected distributions of the reconstructed Higgs (a) and top (b) ($t\to Hc$) mass reweighted by $\ln(1+S/B)$ for all channels. The weights are determined by the signal ($S$) and background ($B$) predictions for each BDT bin. A data luminosity of 100 fb$^{-1}$ and $\BR(t\to Hc)=1\%$ are assumed. The red histograms represent the background, while the black dots represent the sum of the signal and background events. }
\label{fig:mass_postfit}
\end{figure*}

The reconstruction performance of the Higgs mass and the top mass is crucial to establish the observed signal as a true FCNC process. Figure~\ref{fig:mass_postfit} shows an example of the expected mass distributions with an integrated luminosity of 100~fb$^{-1}$, where $\BR(t\to Hc)=1\%$ is assumed. In Fig.~\ref{fig:mass_postfit}, each event entry is weighted by $\ln (1+S/B)$, which is determined by the signal ($S$) and background ($B$) predictions for each BDT bin.

\section{Conclusions}
To conclude, based on a fit taking into account the Higgs mass constraint and the $\tau$ decay kinematics, both the Higgs mass and the top mass in the decay $t \to Hc, H \to\tau\tau$ can be reconstructed with good resolutions. Together with other discriminating variables, a BDT-based analysis shows that $\BR(t\to Hc)$ can be probed down to 0.25\% with a data set of 100~fb$^{-1}$  in LHC Run-2. Thus, the coupling constant $|\lambda_{tcH}|<0.096$ is obtained according to the formula $|\lambda_{tcH}| = 1.92\sqrt{\BR(t\to Hq)}$ in Ref.~\cite{fcnc_atlas}. It is worth noting that the results in this analysis are based on the expected ATLAS performance in the Run-1 period, especially for the $b$-tagging, which is expected to improve in Run-2 due to the ATLAS inner detector upgrade \cite{BTag}. The trigger acceptance is not studied in this analysis, whose impact can be somehow compensated by the fact that a large fraction of Run-2 data is expected to be taken at 14 TeV collision energy. The result can be combined with $t\to Hc, H\to\gamma\gamma/b\bar{b}$ analyses in which the masses of Higgs and top are also reconstructed for discovery, and can be further combined with the tri-lepton searches where these masses are not reconstructed for exclusions. The method introduced here can be applied equally well to the $t\to Hu$ decay process, as no explicit $c$-tagging is used in this work.

\section{Acknowledgements}
The authors are grateful for the help from Hongjian He (who initiated discussion about this channel with us), Ruiqing Xiao, Qing Wang and Ning Zhou. We acknowledge the support from Tsinghua University, Center of High Energy Physics of Tsinghua University and Collaborative Innovation Center of Quantum Matter of China, and the National Thousand Young Talents program (20151710211). L.-G. Xia is supported by the General Financial Grant from the China Postdoctoral Science Foundation (Grant No. 2015M581062).

\end{document}